\documentclass[11pt]{article}\pdfoutput=1
\usepackage{amsmath,amsfonts,amssymb,amscd}
\usepackage{pstricks,pst-node}
\usepackage[small,bf,hang]{caption}
\usepackage{tensor}
\usepackage{slashed}
\usepackage{amsmath}
\usepackage{latexsym,epsfig}
\usepackage{arydshln}
\usepackage{extarrows}
\usepackage[utf8]{inputenc}
\usepackage[linktoc=all,hidelinks]{hyperref}
\usepackage{cite}

\usepackage{amsthm,tikz-cd,accents}

\newcommand\lf{$L_\infty${}}
\DeclareMathOperator{\im}{im}

\newtheorem*{proposition}{Proposition}

\usepackage[makeroom]{cancel}

\def\hybrid{
        \topmargin -20pt
        \oddsidemargin 0pt
        \headheight 0pt \headsep 0pt
        \textwidth 6.25in 
        \textheight 9.5in 
        \marginparwidth .875in
        \parskip 5pt plus 1pt \jot = 1.5ex}

\hybrid

 \csname
@addtoreset\endcsname{equation}{section}

\def\moth{\mathsurround=0pt}
\newdimen\zo \zo=0pt

\def\tick{\leaders\hrule height 0.5ex depth 0pt \hskip 0.5pt}
\def\upboxfill{$\moth \setbox\zo\hbox{\tick}%
  \hskip 3pt\hbox to 0pt{$\tick$\hss}\hrulefill \hbox to 7.5pt{$\tick$\hss}$}

\def\dtick{\leaders\hrule height .34pt depth 0.5ex \hskip 0.5pt}
\def\downboxfill{$\moth \setbox\zo\hbox{\dtick}%
  \hskip 2pt\hbox to 0pt{$\dtick$\hss}\hrulefill \hbox to 2pt{$\dtick$\hss}$}

\def\pd{\partial}
\def\one{{\bf 1}}

\newcommand{\overbar}[1]{\mkern 2mu\overline{\mkern-1mu#1\mkern-1mu}\mkern 1mu}

\def\bec{\begin{center}}
\def\ec{\end{center}}

\def\cC{{\cal C}}

\def\cO{{\cal O}}

\def\cM{{\cal M}}

\def\cO{{\cal O}}

\def\be{\begin{equation}}
\def\ee{\end{equation}}
\def\bea{\begin{eqnarray}}
\def\eea{\end{eqnarray}}
\def\ba{\begin{array}}
\def\ea{\end{array}}

\def\ket#1{|#1\rangle}

\begin{document}

\begin{titlepage}
June 2021 \hfill {HU-EP-21/09}
\rightline{Imperial-TP-2021-CH-01}
\begin{center}
\vskip 1cm
{\Large \bf{Homotopy Transfer and Effective Field Theory II: \\[1.5ex] 
Strings and Double Field Theory
}
}\\
\vskip 1cm

 {\large {Alex S.~Arvanitakis$^{a}$, Olaf Hohm$^{b}$, Chris Hull$^{c}$ and Victor Lekeu$^{d}$ }}
\vskip .5cm

{\it  
      $^a$Theoretische Natuurkunde, Vrije Universiteit Brussel, \\ and the International Solvay Institutes, \\ Pleinlaan 2, B-1050 Brussels, Belgium\\ \ \\}

{\it  $^b$Institute for Physics, Humboldt University Berlin,\\
 Zum Gro\ss en Windkanal 6, D-12489 Berlin, Germany\\ \ \\}

{\it $^c$The Blackett Laboratory, Imperial College London, \\
Prince Consort Road, 
London
SW7 2AZ, 
U.K.\\ \ \\}

{\it $^d$Max-Planck-Institut für Gravitationsphysik (Albert-Einstein-Institut)\\
Am Mühlenberg 1, 14476 Potsdam, Germany\\ \ \\}
 
\vskip .1cm

alex.s.arvanitakis@vub.be, ohohm@physik.hu-berlin.de, c.hull@imperial.ac.uk, victor.lekeu@aei.mpg.de

\vskip 1cm
{\bf Abstract}

\end{center}


\noindent
\begin{narrower}

 We continue our study of effective field theory via homotopy transfer of $L_\infty$-algebras, and apply it to tree-level non-Wilsonian effective actions of the kind discussed by Sen in which the modes integrated out are comparable in mass to the modes that are kept. We focus on the construction of effective actions for string states at fixed levels and in particular on the construction of weakly constrained double field theory. With these examples in mind, we discuss closed string theory on toroidal backgrounds and resolve some subtle issues involving vertex operators, including the proper form of cocycle factors and of the reflector state. This resolves outstanding issues concerning the construction of covariant closed string field theory on toroidal backgrounds. The weakly constrained double field theory is formally obtained from closed string field theory on a toroidal background by integrating out all but the \lq doubly massless' states and homotopy transfer then gives a prescription for determining the theory's vertices and symmetries. We also discuss consistent truncation in the context of homotopy transfer.

\end{narrower}

\end{titlepage}

\tableofcontents


\section{Introduction}

The Wilsonian effective action is obtained by integrating out all heavy degrees of freedom  above a certain mass scale to obtain an effective description of the remaining light modes.
This paper, which is a continuation of \cite{Arvanitakis:2020rrk}, explores more general effective field theories obtained by integrating out a subsector of the degrees of freedom without the requirement that they be heavier than the modes that are kept. 
Our goal here is  to 
explore and  clarify the issues that arise in the construction of such effective theories, following \cite{Sen:2016qap}, 
and then to apply this to string theory 
with a particular focus on the construction of double field theory.

We   use the formulation of classical or tree-level field theories in terms of  $L_{\infty}$-algebras, which encode  the theory's interactions and symmetries according to the general dictionary of  \cite{Hohm:2017pnh}, and using the notion of homotopy transfer explained in our previous 
 paper  \cite{Arvanitakis:2020rrk}. (The case beyond tree-level requires an extension to loop $L_{\infty}$-algebras, which will be 
explored in a subsequent   paper.) 
We argued in \cite{Arvanitakis:2020rrk} that the field theory procedure of integrating out  {a set of} modes at tree-level is described 
algebraically in terms of homotopy transfer; see also the related publications  \cite{Erbin:2020eyc,Koyama:2020qfb}. 
{In particular, homotopy transfer provides an algebraic  construction of the $L_{\infty}$-algebra encoding  the effective theory's interactions and symmetries
from that of the original theory.}
Aspects of  {the} relation between integrating out degrees of freedom and homotopy transfer appear to have been known to experts, 
with instances of this relation being  explored for particular models in e.g.~\cite{Mnev:2006ch,Cattaneo:2008ph,Alexandrov:2007pd}, 
and in the $A_{\infty}$ framework relevant for open string theory in \cite{Kajiura:2001ng,Kajiura:2003ax}, 
but 
to the best of our knowledge it was only in
\cite{Arvanitakis:2020rrk,Erbin:2020eyc,Koyama:2020qfb}
 that the role of homotopy transfer for \textit{general} field theories
requiring $L_{\infty}$-algebras
has been discussed explicitly. Apart from being of conceptual interest, this relation is of importance for applications starting from (classical) closed string field theory, 
which in Zwiebach's covariant  formulation is governed by an $L_{\infty}$-algebra \cite{Zwiebach:1992ie}.

The main example we have in mind is that of the construction of
a double field theory as an effective field theory for closed string theory on a  toroidal background \cite{Siegel:1993th,Siegel:1993bj,Tseytlin:1990nb,Duff:1989tf,Hull:2009mi}.
On a toroidal background, the metric, B-field and dilaton all become {\it double fields} depending on both momentum and winding number and the  question is whether there is a `true double field theory'  for these double fields obtained from integrating out all other degrees of freedom.
Such theories are referred to in the literature as being {\it weakly constrained}, {but so far most of the work in the field has been on the {\it strongly constrained} theory, which
is a small sub-theory that essentially captures only the local supergravity theory and misses much of the stringy physics.
The strongly constrained theory was originally constructed by Siegel \cite{Siegel:1993th,Siegel:1993bj} and further discussed in  \cite{Hull:2009mi}, where its relation to the weakly constrained theory was emphasized.
It would be of great interest to write {a weakly constrained double field theory} explicitly, since it would describe  genuinely stringy effects due to the presence of fields beyond the supergravity sector 
{and would have manifest T-duality invariance}. The construction of such a {theory} was one of the central goals of \cite{Hull:2009mi},
where the theory was constructed to cubic order in the fields, and it was argued that at higher order in the fields such a theory is non-local.
It was shown by Sen that such  general effective field theories can   be obtained from closed string field theory by 
integrating out the  appropriate fields \cite{Sen:2016qap}.

{Integrating out massive fields  typically results in a theory with non-local interactions  for the remaining fields.} 
At energy scales that are small compared to the masses of the fields that are integrated out, one can typically recover a local field theory description. 
Concretely, upon introducing a cut-off at an energy scale $\Lambda$ which is small compared to the mass scale $M$ of the fields that are integrated out, the non-localities are suppressed by factors of $\Lambda/M$ and vanish in the limit $\Lambda/M \to 0$.
In the literature, the term ``effective field theory" is sometimes used for the theory obtained by integrating out a subset of fields, which is typically non-local, and the term is sometimes 
used for the local field theory that arises from this in a limit such as the one discussed above. In this paper, we will use effective field theory for the former, the non-local theory that arises from intergrating out a subset of fields. As we shall discuss below, there are other ways in which a local field theory can be extracted from a non-local effective field theory, and these will be of particular interest. 

In this paper, we only discuss the tree-level effective action. Integrating out fields at tree level amounts to setting each field being integrated out to a solution of its classical field equations. Moreover, we work in perturbation theory here, so that we only consider perturbative classical solutions constructed around a solution of the linearised field equations.

{We emphasize that     homotopy transfer typically leads  to a non-local theory, as expected from integrating out a subset of fields.
We do not mean to imply that non-local theories obtained in this fashion are free from any physical pathologies. 
Rather, we are making the technical point that reasonable  field theory procedures of integrating out fields are precisely captured (at tree level) 
by homotopy transfer, whether or not the effective field theory may be problematic on other grounds. }

 {In this paper} we     explore and clarify the relation between effective field theory and homotopy transfer. 
In sec.~\ref{sec:integratingout} we illustrate this relation in some field theory models by integrating out massive modes, discussing the relation to consistent truncations and some of the issues arising from gauge symmetries.
In integrating out, particular attention must be paid to zero modes, which are states that live in the kernel of the linearised gauge-fixed kinetic operator and hence render the kinetic operator non-invertible.
 In sec.~\ref{sec:zeromodes} we turn to the zero mode issue and examine the effect of zero modes in homotopy transfer generally, and give a criterion for the consistency of their truncation. 
  In section~\ref{sec:zeromodes} we also discuss   the relation between the $L_{\infty}$-algebra encoding the full theory to the $L_{\infty}$-algebra of the effective field theory 
in more detail.

 In the remainder of this paper we focus on the double field theory sector in order to prove
 that 
there is a homotopy transfer from the full closed string field theory to a true or `weakly constrained' double field theory featuring 
massive Kaluza-Klein and winding modes but no other string modes. 
To this end we review in sec.~\ref{sec:stringstorus} some pertinent features of string theory on toroidal backgrounds.

Our main new results are given in sec.~\ref{sec:sft}, \ref{sec:homotopystrings} and \ref{sec:DFT}. 
In sec.~\ref{sec:sft} we resolve the  issue of ``cocycle sign'' insertions  in the reflector state for covariant string field theory \cite{Zwiebach:1992ie} on a torus. (The reflector is essentially the {BV} antibracket {of string field theory}, as we will review.) {The possibility of these insertions reflects an ambiguity in free closed string field theory, which potentially could lead to inconsistencies when interactions are introduced, even at the cubic level \cite{Hata:1986mz,Kugo:1992md,Maeno:1989uc}. In sec.~\ref{sec:sft}, we study vertex operators and find necessary conditions on these signs that fix this ambiguity; we then conjecture that these signs are correct for the fully interacting theory.} This free field theory result is sufficient in order to check the requirements for homotopy transfer in string theory, to which we turn 
in sec.~\ref{sec:homotopystrings}. There we develop  homotopy transfer in closed string theory, considering the low-energy effective action in Minkowski space and the effective action for string modes of fixed level.
In sec.~\ref{sec:DFT},
we establish the existence of a weakly constrained double field theory obtained by homotopy transfer from closed string field theory.
The resulting theory is specified by its $L_{\infty}$ brackets  that can in principle be derived via the homotopy transfer formulas of \cite{Arvanitakis:2020rrk} from the 
$L_{\infty}$ brackets of the closed string field theory on toroidal backgrounds. However, the complexity of the brackets (see e.g.~\cite{Kugo:1992md} for the 3-point vertex in a non-covariant string field theory) make this difficult to do explicitly and we do not attempt this here.


\section{Integrating out massive fields} 
\label{sec:integratingout}

In this section we discuss and review various issues related to integrating out massive fields (at tree-level) in general, i.e., irrespective of whether the remaining states are lighter or not. 
We illustrate this with a simple $U(1)$ Higgs model and give an interpretation in terms of homotopy transfer of the corresponding $L_{\infty}$-algebras.

{
\subsection{Effective field theories}
\label{sec:zeromodesintro}

Consider a field theory in which we wish to integrate out some subset of fields to obtain an effective field theory. This requires inverting the kinetic operator of the fields to be integrated out, in order to have a well-defined propagator to perform the perturbative path integral. Equivalently, integrating out at tree-level is equivalent to solving the equations of motion of the fields to be eliminated and plugging back the solution into the action; such a solution can be found perturbatively whenever the linearised gauge-fixed kinetic operator is invertible.

Let us then start with a short discussion of the possible issues with this inversion. The linearised kinetic operator $\cO$ is not invertible whenever there are solutions to the equation $\cO A = 0$, which we will  refer to as zero-modes. These come in three qualitatively different kinds:
\begin{itemize}
    \item They can be pure gauge modes. Those can be dealt with by the usual field-theoretical gauge-fixing procedure or by homotopy transfer to gauge invariant variables \cite{Chiaffrino:2020akd}. In our context, the Batalin-Vilkovisky field-antifield formalism \cite{Batalin:1981jr,Batalin:1984jr} is particularly well suited for gauge-fixing, given its natural link with $L_\infty$-algebras; see e.g.~\cite[section 5.2]{Arvanitakis:2020rrk} for an explicit example.
    \item There will be zero-modes corresponding to  physical states of mass $m$ with $p^2=-m^2$. These can be eliminated (as is customary in field theory) by Wick rotating to Euclidean space or, equivalently, by introducing Feynman's $i\epsilon$. After the integrating out is performed, one then needs to Wick rotate back to find an effective field theory in Lorentzian signature.
    \item There can be a further finite-dimensional space of zero-modes even in the Euclidean gauge-fixed theory, for example constant massless fields or, for anti-symmetric tensor gauge fields, harmonic forms. These are non-normalisable on non-compact spaces. To deal with these zero-modes, we decompose the field space into the space of zero-modes and its complement, and consider the path integral over non-zero modes only, using the fact that the Euclidean gauge-fixed kinetic operator is invertible on that space. The effect of the remaining zero-modes then depends on the precise theory at hand: the finite-dimensional integral over those zero-modes would give an infinite volume factor if the zero-modes do not appear in the interactions, it would give constraints if they appear linearly as Lagrange multipliers while if they appear non-linearly, they would act as constant auxiliary fields. These issues are discussed in further detail in section \ref{sec:zeromodes}.
\end{itemize}

In inverting the kinetic operator, integrating out a field of mass $m$ will then typically induce non-localities involving $(p^2+m^2)^{-1}$. These are non-singular in Euclidean signature but have a pole in Lorentzian signature. (These could formally be avoided by including a small $i\epsilon$ term, moving the poles off the real axis.)
If the momenta are restricted to the low energy regime in which $|p^2|< \Lambda ^2$ where the cut-off $\Lambda$ is less than the mass of each field that is integrated out, then the poles are avoided and the low-energy effective action is non-singular.
The propagators can then be expanded
as
\be\label{eq:pexpansion}
\frac 1 {p^2+m^2} =\frac 1 { m^2} \left[ 1- \frac {p^2 } { m^2} + \left( \frac {p^2 } { m^2} \right)^2+\dots \right]
\ee
to give  a derivative expansion. In particular, if $|p^2|<<  m ^2$ then the propagator can be approximated by $1/m^2$ and a local low-energy effective action emerges. However, motivated by Sen's discussion in the string theory context \cite{Sen:2016qap}, we will consider the full non-local result in the examples that follow.

}

\subsection{A toy model for the toy model}

Before turning to the Higgs model we briefly illustrate the main point with a  model without gauge symmetries
for two massive scalars $\phi$ and $\varphi$. 
Consider the action 
 \be
    S=\int d^4x\left(\frac{1}{2}\phi\square\phi -\frac{1}{2}M^2\phi^2+\frac{1}{2}\varphi\square \varphi 
    -\frac{1}{2}m^2\varphi^2 + g \phi J(\varphi)  \right)\;, 
 \ee
 where $J(\varphi)$ is an arbitrary (local) function of $\varphi$. Our goal is to integrate out $\phi$ in order to obtain an effective action for $\varphi$. 
It is convenient to rescale $\phi\rightarrow M\phi$ and $g\rightarrow \frac{1}{M} g$, after which the action can be written as 
  \be
  {S}=\int d^4x \left(-\frac{1}{2}\phi{\cal O}\phi+\frac{1}{2}\varphi\square \varphi 
    -\frac{1}{2}m^2\varphi^2 + g \phi J(\varphi)  \right)\;, 
 \ee
where we have defined the operator 
 \be\label{operator}
  {\cal O}\equiv 1-\frac{1}{M^2}\square \;. 
 \ee 
Integrating out $\phi$ at tree-level amounts to solving the field equations for $\phi$ in terms of $\varphi$ and reinserting into the action.  
The field equations for $\phi$ read ${\cal O}\phi=gJ(\varphi)$, and assuming that the 
operator  ${\cal O}$ can be inverted we can solve for $\phi$:  
 \be\label{phiSolution}
  \phi=g{\cal O}^{-1}J(\varphi)\;. 
 \ee
Formally we can write the inverse of (\ref{operator}) as a geometric series{, as in \eqref{eq:pexpansion}}: 
 \be
  {\cal O}^{-1}=1+\frac{1}{M^2}\square+\frac{1}{M^4}\square^2+\frac{1}{M^6}\square^3+\cdots \;. 
 \ee 
Reinsertion of  (\ref{phiSolution})  into the action yields 
\be
  {S}=\int d^4x \left(\frac{1}{2}\varphi\square \varphi 
    -\frac{1}{2}m^2\varphi^2 + \frac{1}{2}g^2 J(\varphi){\cal O}^{-1} J(\varphi)  \right)\;. 
 \ee
We obtain a   non-local action for the scalar $\varphi$ whose mass $m$ need not be smaller 
than the mass $M$ of the scalar $\phi$ we have integrated out. This procedure therefore does \textit{not} comply with the usual Wilsonian paradigm 
according to which only those modes should be integrated out whose mass scale exceeds the typical scale of processes we are interested in.
The Wilsonian picture is recovered in the limit $M^2\rightarrow\infty$ (keeping $m^2$ and the rescaled $g$ finite) for which the action reduces to 
  \be
  {S}\Big|_{M^2\rightarrow\infty} =\int d^4x \left(\frac{1}{2}\varphi\square \varphi 
    -\frac{1}{2}m^2\varphi^2 + \frac{1}{2}g^2 J(\varphi)^2  \right)\;. 
 \ee
The effect of integrating out $\phi$ in this limit is the appearance of a new interaction term for $\varphi$ proportional to $g^2$ that is completely local. 
(For instance, if the original action is cubic integrating out $\phi$ induces a quartic interaction term.) As an  aside we remark that it is known that 
in string field theory massive string modes have to be integrated out, along the lines above, in order to 
produce the higher interaction vertices for massless fields (such as the quartic vertex of Yang-Mills theory, which follows from cubic 
string field theory) \cite{Siegel:1988yz}.

\subsection{A Higgs model}

After these general remarks we now turn to a more realistic model with gauge symmetries 
in which case the homotopy  $L_{\infty}$-algebra interpretation will become more subtle and interesting. 
We consider a Higgs model with $U(1)$ gauge symmetry, with the fields being a $U(1)$ gauge field $A_{\mu}$ and a 
complex scalar $\phi$. The Lagrangian  reads 
 \be
  {\cal L}=- \frac{1}{4} F^{\mu\nu} F_{\mu\nu} - \frac{1}{2} D^{\mu}\phi^* D_{\mu}\phi - V(\phi)\;, 
 \ee
where $F_{\mu\nu}=2\partial_{[\mu} A_{\nu]}$ is the abelian field strength, and the covariant derivatives and scalar potential are 
 \be
  D_{\mu}\phi=\partial_{\mu}\phi-iA_{\mu}\phi\;, \qquad V(\phi)=-\frac{1}{2}\mu \phi^*\phi +\frac{\lambda}{4}(\phi^*\phi)^2\;. 
 \ee
This model is  gauge invariant under $A_\mu \rightarrow A'_\mu = A_\mu + \pd_\mu \Lambda$, $\phi\rightarrow\phi'= e^{i\Lambda}\phi$, $\Lambda\in \mathbb{R}$, or, infinitesimally, under 
 \be\label{U(1)Gauge}
  \delta A_{\mu}=\partial_{\mu}\Lambda\;, \qquad \delta\phi= i\Lambda\phi\;. 
 \ee
It is convenient to split the complex scalar into two real scalars, one of which is gauge invariant, the other pure gauge. Writing 
 \be\label{Phipara}
  \phi =  r  e^{i\varphi}\;, 
 \ee 
the  gauge transformations become 
 \be\label{U(1)Gauge2}
  \delta \varphi=\Lambda\;, \qquad \delta r=0\;, 
 \ee
confirming that $r$ is gauge invariant.  Writing out the covariant derivatives and using the parametrization (\ref{Phipara}) 
the action reduces to 
 \be
   {\cal L}=- \frac{1}{4} F^{\mu\nu} F_{\mu\nu}   - \frac{1}{2}r^2 A^{\mu} A_{\mu} + A_{\mu} j^{\mu} 
   -\frac{1}{2} r^2 \partial^{\mu}\varphi \partial_{\mu}\varphi - \frac{1}{2} \partial^{\mu}r \partial_{\mu} r - V(r)\;, 
 \ee
with current and scalar potential 
\be
 j^{\mu}=r^2 \partial^{\mu}\varphi\;, \qquad V(r) = -\frac{1}{2}\mu r^2 +\frac{\lambda}{4}r^4\;. 
\ee

We will next expand the above action about a (constant) vacuum solution $\langle\phi\rangle = \langle r\rangle =v$ that spontaneously breaks the $U(1)$. 
Specifically, assuming $\mu,\lambda>0$ we can pick a vacuum with $v^2=\frac{\mu}{\lambda}$ that minimizes the potential energy 
and then expand 
 \be
  r = v +\rho\;, \qquad  v=\sqrt{\frac{\mu}{\lambda}}\;, 
 \ee 
where the fluctuation  $\rho$ is the Higgs field. Note that $\varphi$, which does not receive a vacuum expectation value, is treated as a pure perturbation.  
This  yields   the action  
 \be\label{fullHiggsexpanded}
  \begin{split}
   {\cal L} = &\,-\frac{1}{4} F^{\mu\nu} F_{\mu\nu}-\frac{1}{2} v^2 A^{\mu} A_{\mu} + v^2 A_{\mu}\partial^{\mu}\varphi
   -\frac{1}{2} v^2 \partial^{\mu}\varphi \partial_{\mu}\varphi - \frac{1}{2}\partial^{\mu} \rho\partial_{\mu}\rho -\mu \rho^2 \\[0.5ex]
   &\, -v \rho A^{\mu} A_{\mu} + 2v \rho A_{\mu} \partial^{\mu}\varphi - v \rho \partial^{\mu}\varphi \partial_{\mu}\varphi - \lambda v \rho^3\\[0.5ex]
   &\, -\frac{\lambda}{4} \rho^4 -\frac{1}{2} \rho^2 A^{\mu} A_{\mu} +\rho^2 A^{\mu}\partial_{\mu}\varphi - \frac{1}{2} \rho^2 \partial^{\mu}\varphi \partial_{\mu}\varphi\;, 
  \end{split}
 \ee
 where the first line consists of terms quadratic in the fields, the second line has cubic terms and the third line has quartic terms.
At this stage it is customary to pick unitary gauge by setting $\varphi=0$, which of course simplifies the analysis, 
but in order to explore how the integrating out of degrees of freedom interferes with gauge symmetries 
it will be instructive  to keep the gauge redundant formulation 
with  fields $A_{\mu}$, $\varphi$ and $\rho$. 
The physical content can, however, still  be brought out by noting that in terms of 
 \be\label{gaugeinvariantA}
  A_{\mu}':=A_{\mu}-\partial_{\mu}\varphi\;, 
 \ee
 with field strength $F=dA'=dA$,
the action (\ref{fullHiggsexpanded}) can be written as 
 \be\label{fullHiggsexpandedGI}
  \begin{split}
   {\cal L} = &\,-\frac{1}{4} F^{\mu\nu} F_{\mu\nu}-\frac{1}{2}v^2 A'^{\mu} A'_{\mu} - \frac{1}{2}\partial^{\mu} \rho\partial_{\mu}\rho -\mu \rho^2 \\[0.5ex]
   &\, -v \rho A'^{\mu} A'_{\mu}   - \lambda v \rho^3  -\frac{\lambda}{4} \rho^4 -\frac{1}{2} \rho^2 A'^{\mu} A'_{\mu} \;. 
  \end{split}
 \ee
 The combination (\ref{gaugeinvariantA}) is gauge invariant under (\ref{U(1)Gauge}), (\ref{U(1)Gauge2}), as is the Higgs field $\rho$, and so 
this action is manifestly gauge invariant. The free action in turn  displays the propagating degrees of freedom:  a massive gauge boson $A_{\mu}'$ (in the St\"uckelberg formulation 
of Proca theory) and a massive scalar $\rho$ (the Higgs boson).

Our first goal is now to integrate out the massive gauge boson to obtain an effective action for the Higgs boson. 
We start from the quadratic part of the action, which we rewrite as  
 \be\label{freeaction}
  {\cal L}^{(2)}=-\frac{1}{2}A^{\mu} {\cal O}_{\mu}{}^{\nu} A_{\nu} + v^2 A_{\mu}\partial^{\mu}\varphi 
  -\frac{1}{2} v^2 \partial^{\mu}\varphi \partial_{\mu}\varphi - \frac{1}{2}\partial^{\mu} \rho\partial_{\mu}\rho -\mu \rho^2\;,
 \ee
where we defined the operator 
 \be\label{MaxwellOperator}
  {\cal O}_{\mu}{}^{\nu} \equiv  v^2(\delta_{\mu}{}^{\nu}- v^{-2} P_{\mu}{}^{\nu})\;, \qquad 
  P_{\mu}{}^{\nu}\equiv \square \delta_{\mu}{}^{\nu}-\partial_{\mu}\partial^{\nu}\;,
 \ee
in terms of  the Maxwell operator 
that encodes  the Maxwell equations as $P_{\mu}{}^{\nu}A_{\nu}=0$.
The free field equations for $A_{\mu}$ then read 
 \be\label{lowestorderEQA}
  {\cal O}_{\mu}{}^{\nu} A_{\nu}=v^2\partial_{\mu}\varphi\;. 
 \ee
We now assume again invertibility of ${\cal O}$, whose  inverse exists at least formally as a geometric series: 
 \be\label{InverseOMaxwell}
  ({\cal O}^{-1})_{\mu}{}^{\nu} = v^{-2} \delta_{\mu}{}^{\nu} +v^{-4} P_{\mu}{}^{\nu} + v^{-6} P_{\mu}{}^{\rho} P_{\rho}{}^{\nu}+\cdots \;. 
 \ee  
Note, in particular, that it is not an issue that the Maxwell kinetic operator $P_{\mu}{}^{\nu}$ is    \textit{not} invertible due 
to gauge invariance (it is subject to the identity $P_{\mu}{}^{\nu}\partial_{\nu}\chi\equiv 0$). 
The lowest-order equation (\ref{lowestorderEQA}) can now be solved for $A_{\mu}$: 
 \be\label{AExpansion}
  A_{\mu}= v^2 ({\cal O}^{-1})_{\mu}{}^{\nu}\partial_{\nu}\varphi\;. 
 \ee
Due to the identity $P_{\mu}{}^{\nu}\partial_{\nu}\chi\equiv 0$ noted above   we now observe that,   using  (\ref{InverseOMaxwell}),  only the first term 
survives, so that the exact solution actually takes the local form 
 \be\label{lowestAmu}
  A_{\mu}=\partial_{\mu}\varphi\;. 
 \ee
Re-substituting this into (\ref{freeaction}) all $\varphi$-dependent terms cancel, leaving only the  kinetic term for the Higgs field $\rho$. 
The same result would of course follow in the gauge fixed formulation with $\varphi=0$ (or equivalently after the field redefinition to gauge invariant 
variables (\ref{gaugeinvariantA})), for which $A_{\mu}$ (or the gauge invariant $A_{\mu}'$) is set to zero directly. 
It is, however, reassuring to see that the procedure of integrating out massive modes works consistently without the need to fix a gauge. 

We will now show that integrating out $A_{\mu}$ amounts to re-substituting (\ref{lowestAmu}) in the action to all orders, 
as indeed is clear in the gauge-fixed formulation. The exact field equations for $A_{\mu}$ following from (\ref{fullHiggsexpanded}) are given by 
  \be
   {\cal O}_{\mu}{}^{\nu} A_{\nu}= v^2\partial_{\mu}\varphi -2 v \rho A_{\mu} + 2 v \rho\partial_{\mu}\varphi -\rho^2 A_{\mu} + \rho^2\partial_{\mu}\varphi\;. 
  \ee
 We now solve this equation perturbatively by making the  ansatz   
  \be
   A_{\mu}=A_{\mu}^{(1)} + A_{\mu}^{(2)} + A_{\mu}^{(3)}+\cdots 
  \ee 
with  the superscript denoting  the power of fields that are kept (i.e.~$\varphi$ and $\rho$).  
By (\ref{lowestAmu}) we have to lowest order $A_{\mu}^{(1)}=\partial_{\mu}\varphi$. 
Evaluating  the equation  to the next two orders yields 
  \be
  \begin{split}
   {\cal O}_{\mu}{}^{\nu} A^{(2)}_{\nu}&= -2 v \rho A_{\mu}^{(1)} + 2 v \rho \partial_{\mu}\varphi    \;, \\
   {\cal O}_{\mu}{}^{\nu} A^{(3)}_{\nu}&=  -2 v \rho A^{(2)}_{\mu}  -\rho^2 A^{(1)}_{\mu} + \rho^2\partial_{\mu}\varphi\;. 
  \end{split} 
  \ee
Using $A_{\mu}^{(1)}=\partial_{\mu}\varphi$  in the first equation gives  ${\cal O}A^{(2)}=0$ and hence $A^{(2)}=0$. Using this again  in the second equation 
one finds ${\cal O}A^{(3)}=0$ and hence $A^{(3)}=0$. Therefore, (\ref{lowestAmu}) is the exact (perturbative) solution. 
Inserting this into the full action (\ref{fullHiggsexpanded}) one obtains 
   \be\label{intoutA}
  \begin{split}
   {\cal L} =  - \frac{1}{2}\partial^{\mu} \rho\partial_{\mu}\rho -\mu \rho^2 - \lambda v \rho^3  -\frac{\lambda}{4} \rho^4 \;. 
  \end{split}
 \ee
 The claim is that this is the effective tree-level action for the Higgs field alone, i.e.,  the correct 
 action for processes whose external states involve only the Higgs mode $\rho$. This can also 
 be seen directly by inspecting the Lagrangian in the form (\ref{fullHiggsexpandedGI}): Since $A_{\mu}'$ 
appears  only quadratically there are no tree-level diagrams with internal lines for $A_{\mu}'$ and only external  $\rho$ states. 
 Thus, the tree-level effective action for $\rho$ can be  obtained by 
 setting $A_{\mu}'=0$. Note that this is in contrast to integrating out $\rho$, which couples linearly to $A_{\mu}'$, 
 so that $\rho$ can appear in internal lines for tree-level diagrams with  external $A_{\mu}'$ states, 
 and hence  it is inconsistent to set $\rho=0$.

 Let us then turn to the problem of integrating out the Higgs field $\rho$ in order to obtain an effective action 
 for the massive gauge boson alone. 
 Focusing  on the terms up to cubic order  involving the Higgs field (and finally suppressing the prime on $A_{\mu}$) 
 we read off from (\ref{fullHiggsexpandedGI}): 
 \be
  {\cal L}_{\rho} =-\frac{1}{2} \rho{\cal D} \rho - v \rho A^{\mu} A_{\mu} -\lambda v \rho^3+\cdots\,, 
 \ee
where 
 \be
  {\cal D} \equiv  2\mu-\square \;, 
 \ee
and the ellipsis denotes  quadratic terms for $A$ and all quartic couplings. 
The equation of motion for $\rho$ then reads  
 \be\label{DphiHiggs}
  {\cal D}\rho = -v A^{\mu} A_{\mu} -3\lambda v \rho^2 +\cdots \;. 
 \ee
Making a perturbative ansatz as above, writing $\rho=\rho^{(1)}+\rho^{(2)}+\cdots$, this is solved by $\rho^{(1)}=0$ and 
 \be\label{leadingphisolution}
  \rho^{(2)} = -v \,{\cal D}^{-1}(A^{\mu} A_{\mu} )\;.  
 \ee
Here the inverse of ${\cal D}$ 
can again be defined formally via a geometric series: 
 \be\label{InverscalD}
  {\cal D}^{-1} = \frac{1}{2\mu} \left(1+\frac{1}{2\mu} \square +\frac{1}{4\mu^2} \square^2+\cdots\right)\;. 
 \ee
Reinserting (\ref{leadingphisolution}) into the action we obtain the effective action for the massive gauge bosons, 
 \be\label{quarticmassivegaugeEFF}
   {\cal L} = \,-\frac{1}{4} F^{\mu\nu} F_{\mu\nu}-\frac{1}{2}v^2 A^{\mu} A_{\mu} + 
   \frac{1}{2} v^2 \, A^{\nu} A_{\nu}\,{\cal D}^{-1} (A^{\mu} A_{\mu}) + \cdots\;, 
 \ee
where the ellipsis denote terms of quintic and higher order, which indeed will be induced by 
the higher order terms of the solution of (\ref{DphiHiggs}). 
This is the non-local effective action for the massive gauge boson. 
We may also take the limit in which the Higgs mass is send to infinity, keeping $v$ and hence the mass of the gauge boson finite. 
To this end we rescale $A\rightarrow \lambda^{\frac{1}{2}} A$ and ${\cal L}\rightarrow \lambda^{-1}{\cal L}$ to obtain 
 \be\label{EffectiveAAction}
  {\cal L} = -\frac{1}{4} F^{\mu\nu} F_{\mu\nu}-\frac{1}{2}v^2 A^{\mu} A_{\mu} + 
   \frac{1}{2} v^2\lambda \, A^{\nu} A_{\nu}\,{\cal D}^{-1} (A^{\mu} A_{\mu}) + \cdots
 \ee
so that, with (\ref{InverscalD}) and recalling $v^2\lambda=\mu$, we find that in the limit $\mu \to \infty$
 \be
  {\cal L}\Big|_{\mu\rightarrow \infty} = -\frac{1}{4} F^{\mu\nu} F_{\mu\nu}-\frac{1}{2}v^2 A^{\mu} A_{\mu} + 
   \frac{1}{4 } (A^{\mu} A_{\mu})^2 + \cdots \;. 
 \ee
The ellipsis again denote  higher order terms  in $A$, which here start with sixth order terms of the   form $\frac{1}{v^2}(A^{\mu} A_{\mu})^3$. 
This is the conventional (and hence local) Wilsonian effective action in the limit that the mass of the Higgs is much larger than the mass 
of the gauge boson.

Let us summarize the general lessons illustrated by the above analysis for the integrating out of a (massive) field at tree-level. 
If the field to be integrated out does not couple linearly to the remaining fields then we may simply set it to zero. 
At tree-level it is clear that the resulting action captures any processes whose external states involve only the states that have been kept. 
Relatedly, the classical theory thus obtained is a consistent truncation of the original theory in the sense that 
any solution of the truncated theory is a solution of the original theory (with the truncated field set to zero). 
If, on the other hand, the field to be truncated couples linearly to the remaining fields it is not consistent to set it to zero 
since the linear coupling induces a source term in the field equation of the truncated field that depends only on fields that are kept.  
Rather, the field to be truncated has to be integrated out as above, which in general leads to new and non-local interaction terms.

\subsection{Homotopy transfer interpretation} 

We now verify that the homotopy transfer of the $L_{\infty}$-algebra encoding the Higgs model leads to the $L_{\infty}$-algebra 
encoding the effective field theory in which the appropriate massive modes have been integrated out. 
We begin by describing the free theory (\ref{freeaction}) and its gauge symmetries in terms of a chain complex:  
a chain  of vector spaces $X_i$ with maps (abstract differentials) $\partial_i:X_i\rightarrow X_{i-1}$ 
 that square to zero in  that $\partial_{i-1}\circ \partial_i=0$. 
 For the $U(1)$ Higgs model the chain complex reads  
  \be\label{chaimcomplexMaxwell}
\begin{array}{ccccccccccc} X_1 &\xlongrightarrow{\partial_1} &X_0 {}  &\xlongrightarrow{\partial_0}
&X_{-1} &
\xlongrightarrow{\partial_{-1}}& X_{-2}
\\[1.5ex]
\{\Lambda\}& &\{{\cal A}\} & &\{{\cal E}\}&
&\{{\cal G}\}
\end{array}
\ee
where $X_1$ is the space of gauge parameters $\Lambda$, $X_0$ is the space of fields denoted collectively ${\cal A}\equiv (A_{\mu}, \varphi, \rho)$,   $X_{-1} $ is the space of field equations ${\cal E}\equiv (E^{\mu}_{A}, E_{\varphi}, E_{\rho})$ and $X_{-2} $ is the space of Noether identities.
The differential $\partial_1$ mapping gauge parameters to fields is given by 
 \be\label{LambdaDIFF}
  \partial_1(\Lambda) =  \begin{pmatrix} \partial_{\mu}\Lambda \\ \Lambda \\ 0 \end{pmatrix} \;, 
 \ee
so that the linearized gauge transformations are encoded in  $\delta {\cal A}=\partial_1(\Lambda)$. 
The differential $\partial_0$ mapping fields to field equations is given by 
 \be\label{partialFIELDS}
  \partial_0({\cal A})=\partial_0 \begin{pmatrix} A_{\mu} \\ \varphi \\ \rho \end{pmatrix} 
  = \begin{pmatrix} -{\cal O}_{\mu}{}^{\nu} A_{\nu} + v^2 \partial_{\mu}\varphi \\ v^2 (\square \varphi -\partial_{\mu}A^{\mu}) \\ (\square -2\mu) \rho \end{pmatrix}\,,
 \ee
so that the linearized field equations are encoded in $\partial_0({\cal A})=0$. 
The nilpotency of the differential, $\partial_0\circ \partial_{1}=0$, then encodes the linearized gauge invariance of the free theory. 
Finally, the differential on the space of field equations reads 
 \be
  \partial_{-1}({\cal E}) = \partial^{\mu}(E_A)_{\mu}-E_{\varphi}\;, 
 \ee
and the Noether identity following from gauge invariance is encoded in $\partial_{-1}(\partial_{0}{\cal A})\equiv 0$.

The above data defining the free theory is sufficient to establish the existence of homotopy transfer, but in order to compute 
the non-linear effective action from the homotopy transfer theorem explicitly we need to define the full $L_{\infty}$-algebra 
on the chain complex (\ref{chaimcomplexMaxwell}). The $L_{\infty}$ brackets 
can be determined by demanding that  the field equations take the $L_\infty$ Maurer-Cartan form 
 \be\label{MaurerCartanEOM}
  0= \partial{\cal A}  + \frac{1}{2}\big[{\cal A},{\cal A}\big] + \frac{1}{3!}\big[{\cal A},{\cal A},{\cal A}\big] + \dots\;, 
 \ee 
 which in this example stops with the three-bracket.
Similarly, we can write  the action in terms of the $L_{\infty}$ brackets by defining the inner product between fields and field equations as 
$\langle {\cal A}, {\cal E}\rangle=\int d^4x (A_{\mu}E_{A}^{\mu}+\varphi E_{\varphi}+
\rho E_{\rho})$, 
 \be\label{generalActionasLinfty}
  S = \frac{1}{2}\big\langle {\cal A},\partial{\cal A}\big\rangle +\frac{1}{3!}\big\langle {\cal A}, \big[{\cal A}, {\cal A}\big]\big\rangle +\frac{1}{4!}\big\langle {\cal A}, 
  \big[{\cal A}, {\cal A}, {\cal A}\big]\big\rangle\;. 
 \ee
For  the free (quadratic)  action one may verify using (\ref{partialFIELDS}) that this indeed reproduces (\ref{freeaction}). 
The cubic terms in turn  are defined by the 2-bracket 
 \be\label{2-bracketHiggs}
 \begin{split}
  \big[{\cal A},{\cal A}\big] &= \begin{pmatrix} -4 v \rho A'_{\mu}  \\ 
  -4 v \partial^{\mu}\rho A_{\mu}' - 4 v \rho\partial^{\mu}A_{\mu}' \\ 
  -2 v A'^{\mu} A'_{\mu} - 3! \lambda v \rho^2 \end{pmatrix} \;, 
 \end{split} 
 \ee  
where we use the short-hand notation  $A_{\mu}'=A_{\mu}-\partial_{\mu}\varphi$, and 
the quartic terms are defined by the 3-bracket 
  \be\label{3-bracketHiggs}
 \begin{split}
   \big[{\cal A},{\cal A},{\cal A}\big] &=\begin{pmatrix} -3! \rho^2 A'_{\mu} \\ 
   -3! \rho^2 \partial_{\mu}A'^{\mu}-12\rho\partial_{\mu}\rho A'^{\mu}\\
   -3!\lambda \rho^3 -3! \rho A'^{\mu} A'_{\mu} \end{pmatrix}\;. 
 \end{split} 
 \ee  
Let us note that even though the match with the field equations only determines 
the $L_{\infty}$ brackets on fields for diagonal arguments (all arguments being equal), 
the general brackets may always be reconstructed from 
polarization identities like \cite{Hohm:2017pnh}
 \be\label{polarizationnnn}
  \big[{\cal A}_1,{\cal A}_2\big] = \frac{1}{2}\left(\big[{\cal A}_1+{\cal A}_2 ,{\cal A}_1 + {\cal A}_2\big] - \big[{\cal A}_1,{\cal A}_1\big]
  - \big[{\cal A}_2,{\cal A}_2\big]\right) \;. 
 \ee 
Since the gauge symmetry is abelian there are no higher $L_{\infty}$ brackets (i.e.~with two or more arguments) 
mixing gauge parameters with fields or fields with field equations.

Having defined the $L_{\infty}$-algebra encoding the $U(1)$ Higgs model let us now discuss the homotopy transfer. 
We begin with the case corresponding to integrating out the massive gauge boson. The homotopy transfer maps the full 
$L_{\infty}$-algebra to the $L_{\infty}$-algebra on the subset of fields that are kept, which here is the Higgs boson $\rho$.  
Formally, this is encoded in the projection and inclusion maps (on the space of fields) 
 \be\label{projectortorho}
  p\begin{pmatrix} A_{\mu} \\ \varphi \\ \rho \end{pmatrix}  = \rho \,, \qquad
  \iota(\rho) = \begin{pmatrix} 0 \\ 0\\ \rho \end{pmatrix}\;, 
 \ee
 which obey $p\iota={\bf 1}$ acting on $\rho$. However, $\iota p$ does not equal the identity; rather, 
homotopy transfer requires that  there are degree $+1$ homotopy maps $h_{-1}:X_{-1}\rightarrow X_{0}$ 
and $h_{0}: X_0\rightarrow X_{1}$ so that  
 \be\label{homotopyRel}
  \iota \circ p = {\bf 1}_0 + \partial_1\circ  h_0 + h_{-1}\circ \partial_0\;, 
 \ee
where the subscripts display the spaces on which these maps act.  In order to find the homotopy maps 
we evaluate $\iota p-{\bf 1}$ acting on fields: 
 \be
  (\iota p-{\bf 1})\begin{pmatrix} A_{\mu} \\ \varphi \\ \rho \end{pmatrix} = \begin{pmatrix} -A_{\mu} \\ -\varphi \\ 0 \end{pmatrix}\;. 
 \ee 
The homotopy relation  (\ref{homotopyRel}) requires this to be equal to  
 \be
 \label{homrell}
  \partial_1 \circ h_0 \begin{pmatrix} A_{\mu} \\ \varphi \\ \rho \end{pmatrix} + h_{-1}\circ \partial_0 \begin{pmatrix} A_{\mu} \\ \varphi \\ \rho \end{pmatrix} = 
  \begin{pmatrix} \partial_{\mu}h_0({\cal A}) \\ h_0({\cal A}) \\ 0 \end{pmatrix} 
  + h_{-1} \begin{pmatrix} -{\cal O}_{\mu}{}^{\nu} A_{\nu} + v^2 \partial_{\mu}\varphi \\ v^2 (\square \varphi -\partial_{\mu}A^{\mu}) \\ (\square -2\mu) \rho \end{pmatrix}\;. 
 \ee
We claim that this equality holds  for 
 \be\label{homotopyMAPPS}
  h_{-1}({\cal E}) \equiv \begin{pmatrix} ({\cal O}^{-1})_{\mu}{}^{\nu} ({E}_A)_{\nu} \\ 0 \\0 \end{pmatrix}\;, \qquad
  h_0({\cal A}) =h_0
  \begin{pmatrix} A_{\mu} \\ \varphi \\ \rho \end{pmatrix} = 
   -\varphi\;. 
 \ee
To check this, we note that (\ref{InverseOMaxwell}) implies
$({\cal O}^{-1})_{\mu}{}^{\nu} v^2 \partial_{\nu}\varphi= \partial_{\mu}\varphi$
so that  \be\label{Oivsersecomp}
  ({\cal O}^{-1})_{\mu}{}^{\nu}(-{\cal O}_{\nu}{}^{\rho} A_{\rho} + v^2 \partial_{\nu}\varphi) = -A_{\mu}+\partial_{\mu}\varphi\;, 
 \ee
 giving the first component of (\ref{homrell}).
We next complete the definition of the projector to the entire chain complex and verify the homotopy relation. 
On gauge parameters we set $p(\Lambda)=0$ (as it should be since the projected theory has no gauge symmetries left), 
and so the homotopy relation requires 
 \be
  (\iota p-{\bf 1})(\Lambda)=-\Lambda =h_0(\partial_1\Lambda)\;. 
 \ee
If follows immediately with (\ref{LambdaDIFF}) and (\ref{homotopyMAPPS}) that this relation is satisfied. 
Finally, defining the projector on the space of field equations 
\be 
p({\cal E})=E_{\rho}
\ee
analogously to (\ref{projectortorho}), i.e., picking out only $E_{\rho}$, 
it is straightforward to verify that the homotopy relation holds provided we define the new homotopy map 
 \be
  h_{-2}({\cal G})=\begin{pmatrix} 0  \\ {\cal G} \\ 0 \end{pmatrix}
 \ee
from the space of Noether identities $X_{-2}$ to the space of field equations $X_{-1}$.  
We note, in particular, that it was crucial to introduce the space of Noether identities in order for the homotopy transfer to work consistently. 
However, this homotopy map will not be needed explicitly in what follows.\footnote{As a side remark we note that 
the detailed formulation of the homotopy transfer is not unique. For instance, we can truncate only $A_{\mu}$ and keep $\varphi$ 
but modify the inclusion map: 
 \be
    p\begin{pmatrix} A_{\mu} \\ \varphi \\ \rho \end{pmatrix}  = \begin{pmatrix} \varphi \\ \rho \end{pmatrix} \,, \qquad
  \iota\begin{pmatrix} \varphi \\ \rho \end{pmatrix} = \begin{pmatrix} \partial_{\mu}\varphi \\ \varphi \\ \rho \end{pmatrix}\;. 
 \ee 
The homotopy relations are then satisfied with only 
the first homotopy map in (\ref{homotopyMAPPS}), as follows with the same computation (\ref{Oivsersecomp}).}

Let us now discuss the transported $L_{\infty}$-brackets on the projected space of the effective field theory. 
In order to determine the 2-bracket one uses the inclusion map to lift the arguments to the original space, 
then takes the 2-bracket there and finally projects back to the space of the effective theory. 
The first step yields with (\ref{2-bracketHiggs}) and the inclusion (\ref{projectortorho}) 
 \be\label{includedbracket}
  [\iota(\rho),\iota(\rho)]= \begin{pmatrix} 0 \\  0 \\ - 3! \lambda v \rho^2 \end{pmatrix} \;. 
 \ee 
Thus, the transported 2-bracket reads 
 \be
  [\rho,\rho] = p( [\iota(\rho),\iota(\rho)]) = - 3! \lambda v \rho^2\;,  
 \ee 
 where by a slight abuse of notation we denote by $[\cdot,\cdot]$ also the 2-bracket on the projected space.  
 The upshot of all this that the cubic couplings are simply obtained by setting $A_{\mu}'=0$. 
 Similarly, for the 3-bracket one finds \cite{Arvanitakis:2020rrk}
  \be\label{transport3BRacket}
  \begin{split}
   [\rho,\rho,\rho]&=p([\iota (\rho),\iota(\rho),\iota(\rho)]) + 3\,p([h([\iota(\rho),\iota(\rho)]), \iota(\rho)])\\
   &=p([\iota (\rho),\iota(\rho),\iota(\rho)]) \\
   &= -3!\lambda \rho^3\;, 
  \end{split}
  \ee
using that $h_{-1}$  (c.f.~(\ref{homotopyMAPPS})) acting on  (\ref{includedbracket}) gives zero. Thus, homotopy transfer tells us that 
the action is obtained by setting $A_{\mu}'=0$, in  agreement with our field theory analysis . 

\medskip

We next turn to the case of integrating out the massive Higgs boson. Since above we have seen
that, as was to be expected,  the $L_{\infty}$ formulation of integrating out fields works well in the presence of gauge symmetries we can now simplify  
by eliminating the gauge redundancy, say by fixing a gauge or by passing over to gauge invariant variables. 
In fact,  the passing over to gauge invariant variables can also be interpreted as 
homotopy transfer  \cite{Chiaffrino:2020akd}.  In the present case this homotopy transfer is defined by the projection and inclusion  maps 
 \be
  p\begin{pmatrix} A_{\mu} \\ \varphi \\ \rho \end{pmatrix}  = \begin{pmatrix} A_{\mu}-\partial_{\mu}\varphi \\ \rho \end{pmatrix}\;, \qquad
  \iota\begin{pmatrix} A_{\mu} \\ \rho \end{pmatrix} = \begin{pmatrix} A_{\mu} \\ 0 \\ \rho \end{pmatrix}\;. 
 \ee
This satisfies the homotopy relation provided we define the homotopy map as in the second equation of (\ref{homotopyMAPPS}), 
i.e., $h({\cal A})=-\varphi$. The homotopy transfer  then yields the full theory written in the gauge invariant form 
(\ref{fullHiggsexpandedGI}). The upshot is that the chain complex has been reduced to two terms, the space of fields and the space of 
field equations, with the only differential  acting between them being
 \be
  \partial{\cal A}= \partial\begin{pmatrix} A \\ \rho \end{pmatrix} = \begin{pmatrix} \partial_{\nu}F^{\nu\mu}+ v^2 A^{\mu}  \\ -{\cal D}\rho \end{pmatrix}\;, 
 \ee
where we recall  the notation ${\cal D} =  2\mu-\square$.  The $L_{\infty}$ brackets on this reduced chain complex 
follow in the obvious fashion from (\ref{2-bracketHiggs}), (\ref{3-bracketHiggs}), e.g., the  2-bracket reads  
 \be\label{nondiagonal2bracket}
  \big[{\cal A}_1,{\cal A}_2\big] = \begin{pmatrix} -2 v (\rho_1 A_{2 \mu}+\rho_2 A_{1 \mu}) \\ -2v A_1^{\mu}A_{2 \mu} -3! \lambda v \rho_1 \rho_2 \end{pmatrix} \;, 
 \ee 
which we wrote for general (non-diagonal) arguments using the polarization identity (\ref{polarizationnnn}).

Let us then inspect the homotopy transfer that corresponds  to integrating out the Higgs boson, starting from 
this reduced $L_{\infty}$-algebra.   The projection and inclusion are 
 \be
  p \begin{pmatrix} A \\ \rho \end{pmatrix} = A\;, \qquad \iota(A) = \begin{pmatrix} A \\ 0 \end{pmatrix}\;, 
 \ee 
and the homotopy relations are obeyed upon choosing the following homotopy map 
 \be
  h({\cal E}) = \begin{pmatrix} 0 \\ {\cal D}^{-1}{E}_{\rho}\end{pmatrix} \;, 
 \ee
which follows as above. 
The transported 2-bracket on the projected space of only massive gauge bosons $A$ then becomes 
 \be
  \big[A, A\big] = p\big([\iota(A),\iota(A)]) = p\left( \left[\begin{pmatrix} A \\ 0 \end{pmatrix},\begin{pmatrix} A \\ 0 \end{pmatrix}\right]\right)
  = p  \begin{pmatrix} 0 \\ -2 v A^{\mu} A_{\mu} \end{pmatrix}  = 0 \;, 
 \ee
where we used (\ref{nondiagonal2bracket}).  
We find that the transported 2-bracket vanishes. This means that there are no cubic couplings, which is in agreement 
with our field theory result (\ref{EffectiveAAction}). 
Next, we can compute the transported 3-bracket as in (\ref{transport3BRacket}), using (\ref{nondiagonal2bracket}) and that the original 3-bracket (\ref{3-bracketHiggs}) 
vanishes upon setting $\rho=0$: 
 \be\label{transportedBRacket}
  \begin{split}
   \big[A,A,A\big]
   &=3\,p([h([\iota(A),\iota(A)]), \iota(A)])\\
   &=3\,p\left(\left[h\begin{pmatrix} 0 \\ -2 v A^{\mu} A_{\mu} \end{pmatrix}, \begin{pmatrix} A \\ 0 \end{pmatrix}\right]\right)\\
   &=3\,p\left(\left[\begin{pmatrix} 0 \\ -2 v {\cal D}^{-1}(A^{\mu} A_{\mu}) \end{pmatrix}, \begin{pmatrix} A \\ 0 \end{pmatrix}\right]\right)\\
   &=3\,p \begin{pmatrix} 4 v^2 {\cal D}^{-1}(A^{\nu} A_{\nu})\, A_{\mu} \\ 0  \end{pmatrix} \\
   &= 12\, v^2 {\cal D}^{-1}(A^{\nu} A_{\nu}) A_{\mu}\;. 
  \end{split}
 \ee
This 3-bracket in turn  determines the quartic action according to (\ref{generalActionasLinfty}), 
 \be
 \begin{split}
  {\cal L}_{\rm quartic} = \frac{1}{4!} \langle A, [A,A,A]\rangle 
 = \frac{1}{2} v^2 A^{\mu} A_{\mu} {\cal D}^{-1}(A^{\nu} A_{\nu})\;. 
 \end{split} 
 \ee
This agrees with the quartic couplings induced by the standard field theory procedure of integrating out the Higgs, c.f.~(\ref{quarticmassivegaugeEFF}). 
We have thus confirmed that integrating out massive fields in field theory (be it in presence of gauge redundancies or not) is completely captured by 
the algebraic procedure of homotopy transfer.

\section{Zero modes and homotopy transfer}\label{sec:zeromodes}

\subsection{General discussion}
\label{sec:zeromodesgeneral}

{In this section we briefly discuss the issue of zero-modes in the $L_\infty$ language, and the consistency conditions that a truncation of these zero modes must obey in order to lead to a consistent field theory formulated in terms of an $L_{\infty}$-algebra. As discussed in section \ref{sec:zeromodesintro},}
the issue of zero modes of `kinetic operators' of fields to be integrated out (such as ${\cal O}$ appearing in the previous section) arises as follows: in order to define the homotopy maps the inverse of ${\cal O}$ is needed, but if ${\cal O}$ has a non-trivial kernel then the inverse does not exist.

In the $L_\infty$ language, recall that the free equations of motion read $\partial A = 0$ for $A \in X_0$, with pure gauge solutions $A = \pd \Lambda$ for $\Lambda \in X_{-1}$. Hence, pure gauge solutions are trivial in the homology $H(X)$ of the 1-bracket $\pd$, and the non-trivial elements of $H(X)$ at degree zero are exactly the non-trivial solutions of ${\cal O}A=0$ modulo gauge transformations. {These include both the on-shell states and the constant modes discussed in section \ref{sec:zeromodesintro}.} Therefore, {in Lorentzian signature} the homology $H(X)$ encodes the spectrum of physical degrees of freedom.

In order to understand the significance of this observation we recall that 
homotopy transfer has the characteristic property of \emph{preserving the homology} of the underlying chain complex. This means that two theories related by homotopy transfer should 
have the same homology and hence the same physical spectrum.
As the original theory and effective field theory in general have different spectra, the relation between them cannot be precisely that of homotopy transfer.
{This is also true in Euclidean signature in general, where despite the absence of physical on-shell states there could still be zero-modes of the third kind discussed in section \ref{sec:zeromodesintro}, i.e.~constant or harmonic fields.} We now turn to discuss the precise relationship.

We introduce a projector $P:X\to X$ on $X$ that, when restricted to $X_0$, annihilates the fields that are to be integrated out. We require that it commutes with $\pd$ and this will then restrict the action of $P$ on the other spaces $X_i$ to also project out the associated gauge parameters, equations of motion etc., as was seen in the examples of the previous section. This may not fix $P$ uniquely, but for the moment we take any $P$ satisfying these requirements. The complex $X$ then splits into a direct sum
\begin{equation}
    X = PX \oplus (\one-P) X\, ,
\end{equation}
and the homology $H(X)$ splits similarly,
\begin{equation}
    H(X) = H(PX) \oplus H\left( (\one-P) X \right)\, .
\end{equation}
Consider now the case in which the homology of the sector to be integrated out is non-trivial, $H\left( (\one-P) X \right)\neq 0$. Then, $X$ and $PX$ have different homologies, so there exists no homotopy map $h:X\to X$ satisfying the homotopy relation
$P=\one+h\pd + \pd h$. 
Therefore, homotopy transfer cannot be used to produce an \lf-algebra structure on the subspace $PX$ of effective degrees of freedom.

To construct an effective field theory on $PX$ through homotopy transfer, we look for a way to eliminate the homology $H((\one-P)X)$ in a consistent way. In field theory language, this requires that the truncation of zero modes is a ``consistent truncation'' in the usual physical sense: solutions of the truncated theory must also be solutions of the full theory. This criterion has a natural analogue in the language of $L_\infty$-algebras.

Let us then make a small detour and discuss this natural consistency criterion in the \lf-algebra language. Let $(Z,\pd_Z, \dots)$ be an $L_\infty$-algebra and $Y$ a subcomplex of $Z$. We call $Y$ a \emph{consistent $L_\infty$ truncation} of $Z$ if two criteria are met:
\begin{enumerate}
    \item $Y$ should carry an $L_\infty$ structure; and
    \item there exists an injective morphism of $L_\infty$-algebras $E: Y \to Z$.
\end{enumerate}
An injective $L_\infty$ morphism {takes} the form
\be
E = \iota + \cdots\,,
\ee
where the linear map $\iota: Y \to Z$ is an inclusion of $Y$ in $Z$, and the omitted terms are multilinear maps from $Y$ to $Z$ (see e.g.~\cite[section 2.1]{Arvanitakis:2020rrk}). 
$Y$ is {then} a subspace of $Z$ but in general is not a subalgebra of $Z$, as the brackets will in general be different from the ones that would be inherited from $Z$. However in the special case in which the morphism is the inclusion map, $E=\iota$, $Y$ is a subalgebra of $Z$. Indeed, when  $E=\iota$ the condition that $E$ is a morphism is equivalent to the closure of the \lf-brackets of $Z$ on the subspace $Y$. In the field theory language, the first requirement is just the fact that the truncated theory is consistent by itself, and the second requirement is the $L_\infty$ translation of what we refer to as  
``consistent truncation'' in the following. It should be emphasized that this use of consistent truncation may deviate from a direct interpretation that views a truncation 
as setting to zero a subset of fields or modes. Here we allow for a more general interpretation in which such modes are not necessarily set to zero but may instead be 
(generally non-linear) functions of the modes that are kept, such that any solution of the ``truncated" theory uplifts to a solution of the full theory. 
(While perhaps somewhat unconventional this use of consistent truncation is standard in the context of Kaluza-Klein truncations, for instance.) Indeed, a morphism  $E$ of \lf-algebras carries solutions to the $L_\infty$ Maurer-Cartan equation \eqref{MaurerCartanEOM} (modulo gauge transformations) 
for the truncated algebra $Y$ to solutions of the $L_\infty$ Maurer-Cartan equation for the original algebra $Z$ (see e.g.~\cite[Theorem 7.8]{doubek2007deformation}). This is easy to see in the case of no interactions, where the only brackets are the 1-brackets $\pd_Z$ and $\pd_Y$: then $E = \iota$ and the morphism condition reads
\be
\iota \circ \pd_Y = \pd_Z \circ \iota\,, 
\ee
from which we see that if $\psi \in Y_0$ solves $\pd_Y \psi = 0$, then $\iota \psi \in Z_0$ solves $\pd_Z (\iota \psi) = 0$.

{We now return to the construction of effective theories. Starting from the $L_{\infty}$-algebra on $X$ for the full theory, we will sketch two ways of eliminating the homology of $(\one - P) X$ to construct an effective theory on $PX$. We begin by identifying the homology $H(X)$ with a linear subspace of $X$. This requires a non-canonical choice of representatives, like a gauge choice, in order to define an injective linear map $\mathfrak{i}: H(X) \rightarrow X$ satisfying $\mathfrak{p} \circ \mathfrak{i} = \one_{H(X)}$, where $\mathfrak{p}(x) = [x]$ is the canonical map from $X$ to $H(X)$ mapping $x$ to its equivalence class $[x]$. {This condition simply says that a homology class $c\in H(X)$ is mapped to an element $\mathfrak{i}(c) \in X$ that indeed belongs to the class $c$.} This then defines a projector
\be
\Pi = \mathfrak{i} \circ \mathfrak{p}\, ,
\ee
such that we have the following identification between $H(X)$ and the linear subspace
\be \label{eq:identifyhomology}
\Pi X\cong H(X)\,.
\ee
{The projector property $\Pi^2 = \Pi$ follows from $\mathfrak{p} \circ \mathfrak{i} = \one_{H(X)}$.}
The injection $\mathfrak{i}$ is a \emph{splitting} of the exact sequence
\be
0\to\im \pd\to X\overset{\mathfrak{p}}{\to} H(X)\to 0\, .
\ee
The projector $\Pi$ yields the direct sum $X=\Pi X\oplus (\one-\Pi)X$. It satisfies $\pd \Pi = \Pi \pd = 0$,\footnote{The first of those follows from $\partial \circ \mathfrak{i} = 0$, which holds since each representative of a homology class is closed. The second follows from $\mathfrak{i}\circ\mathfrak{p} \circ \pd(x) = \mathfrak{i} [\pd x] = \mathfrak{i} [0] = 0$.} and we will also assume $\Pi$ commutes with $P$.
}

{
We now present our first way of obtaining an $L_\infty$ structure on $PX$. Our first step retains the homology of $(\one - P) X$. We define the projector
\be\label{Phat}
\widehat{P} = P + (\one-P)\Pi\, .
\ee
The resulting subspace $\widehat{P}X$ is bigger than $PX$, and we have
\begin{equation}
    H(\widehat{P} X) = H(X)\, .
\end{equation}
Therefore, homotopy transfer from $X$ to $\widehat{P}X$ is possible \cite[section 5.1]{Arvanitakis:2020rrk}, and this corresponds to integrating out the non-zero modes in $(\one-P) X$ only. As a second step, it is necessary to check whether the truncation from $\widehat{P}X$ to $PX$ is consistent (in the sense described above with $Z = \widehat{P}X$ and $Y = P X$). If it is consistent, this leads to an $L_\infty$-algebra structure on $PX$. If it is not, we only obtain an $L_\infty$ structure on $\widehat{P}X$. While the first step is always possible, this second step (the existence of a consistent truncation) must be analysed on a case-by-case basis{: in general, the zero-modes of the fields to be integrated out must be kept and can lead to the various effects discussed in section \ref{sec:zeromodesintro}}. This is the strategy that we employ in sections \ref{sec:homotopystrings} and \ref{sec:rforms}.

In our second way of obtaining an $L_\infty$ structure on $PX$, we first attempt to truncate the homology of $(\one - P) X$. The relevant projector is
\begin{equation}
    \widetilde{P} = \one - \Pi (\one - P) \, ,
\end{equation}
which only projects out $H\left( (\one - P) X \right)$, and we have
\begin{equation}
    H(\widetilde{P} X) = H(PX)\, .
\end{equation}
Provided this is a consistent $L_\infty$ truncation of $X$ (in the sense described above with $Z = X$ and $Y = \widetilde{P} X$), one can now use homotopy transfer from $\widetilde{P} X$ to $P X$ to construct an effective theory on $PX$.}

{Let us reiterate that this issue of zero-modes is not specific to the $L_\infty$ language but already exists in the field theory setting.
In the following subsection we will illustrate this with an example in which the zero modes can be separated explicitly, both in field theory and 
in the $L_\infty$ formulation.}

\subsection{Antisymmetric tensor gauge fields on a compact Riemannian manifold}
\label{sec:rforms}

An illustrative example is provided by $r$-form gauge fields on a compact Riemannian manifold without boundary  ${\mathcal{M}}$, where the zero-modes are given by harmonic forms.

\paragraph{Short review of Hodge theory.} Let us start with a short review of relevant facts in Hodge theory in order to be self-contained and fix our notation (see e.g.~\cite[section 7.9]{Nakahara:2003nw} for an accessible introduction). The standard
inner product for $r$-forms is
\begin{eqnarray} \label{eq:innerforms}
( \alpha, \beta ) \equiv \int_{\mathcal{M}} \alpha \wedge ( \star \beta )
\end{eqnarray}
and it is positive-definite in Euclidean signature, i.e.~$(\alpha, \alpha) \geq 0$ for all $\alpha$ in $\Omega^r(\cM)$ with equality iff $\alpha = 0$.
The \emph{adjoint} $d^\dagger$ to $d$ is defined by 
\begin{eqnarray}
( d \beta , \alpha )  = ( \beta , d^\dagger \alpha )\, .
\end{eqnarray}
It reads explicitly $d^\dagger = (-1)^{mr + r + 1} \star d \,\star$ when acting on $r$-forms, with $m = \dim \cM$, and satisfies $(d^\dagger)^2 = 0$ just like $d$. The Laplacian is then
\be \label{eq:lapdef}
\triangle =  d d^\dagger + d^\dagger d 
\ee
and a form $\omega$ is harmonic if 
\be\label{Harmonicc}
\triangle
\omega=0\, .
\ee
The set of harmonic forms is denoted by $\mathrm{Harm}^r(\cM)$. In Euclidean signature, a form is harmonic if and only if it is both closed, $d\omega = 0$, and co-closed, $d^\dagger \omega = 0$. One direction follows immediately from (\ref{eq:lapdef}), and the other  follows from the identity
\begin{equation}
    (\omega, \triangle \omega) = (d\omega, d\omega) + (d^\dagger \omega, d^\dagger \omega)
\end{equation}
together with the positive-definiteness of the inner product.

The Hodge decomposition theorem states that the vector space $\Omega^r(\cM)$ of $r$-forms can be decomposed as
\be
\Omega^r(\cM) = d \Omega^r(\cM) \oplus d^\dagger \Omega^r(\cM) \oplus \mathrm{Harm}^r(\cM)\, ,
\ee
i.e.~any $r$-form $A$ can be written uniquely as the sum of an exact form, a coexact form and a harmonic form:
\be \label{eq:hodgedec}
A = d \Lambda + d^\dagger \beta + A_0
\ee
for some $\Lambda \in \Omega^{r-1}(\cM)$, $\beta \in \Omega^{r+1}(\cM)$ and $A_0 \in \mathrm{Harm}^r(\cM)$. Here, `uniquely' means that the $r$-forms $d \Lambda$ and $d^\dagger \beta$ are unique; this is of course not the case for $\Lambda$ and $\beta$ themselves, which are only defined up to an exact (resp.~coexact) form. This decomposition is orthogonal with respect to the inner product \eqref{eq:innerforms}.

It turns out that the vector space $\mathrm{Harm}^r(\cM)$ is in fact finite-dimensional and the dimension of this space, the number of linearly independent harmonic $r$-forms, is exactly the $r$'th Betti number $b_r$. We shall restrict ourselves to cases in which this is non-zero. This is because each de Rham cohomology class has a unique harmonic representative, so that
\be
\mathrm{Harm}^r(\cM) \cong H^r(\cM)
\ee
(Hodge's theorem) and in particular $\dim \mathrm{Harm}^r(\cM) = \dim H^r(\cM) \equiv b_r$.

We use the notation $\overline{ \mathrm{Harm}^r(\mathcal{M}) } = d \Omega^r(\cM) \oplus d^\dagger \Omega^r(\cM)$ for the complement of $\mathrm{Harm}^r(\mathcal{M})$, which is infinite-dimensional. Using the Hodge decomposition \eqref{eq:hodgedec}, one sees that the image of the Laplacian is contained in $\overline{ \mathrm{Harm}^r(\mathcal{M}) }$, since $\triangle A = d (d^\dagger A) + d^\dagger(dA)$ by \eqref{eq:lapdef}. The restriction of the Laplacian to this space is then well-defined, and we denote it by
\begin{equation}
    \triangle' : \overline{ \mathrm{Harm}^r(\mathcal{M}) } \rightarrow \overline{ \mathrm{Harm}^r(\mathcal{M}) }\, .
\end{equation}
While the full Laplacian $\triangle : \Omega^r(\cM) \rightarrow \Omega^r(\cM)$ is in general not invertible due to the existence of harmonic forms, the operator $\triangle'$ \emph{is} invertible.

We now use the operator $( \triangle ' )^{-1}$ to realise the decomposition \eqref{eq:hodgedec} more explicitly. Let us call $\Pi$ the projector onto harmonic $r$-forms,
\begin{equation}
  \Pi \;:\; \Omega^r(\mathcal{M})  \rightarrow  \mathrm{Harm}^r(\mathcal{M}) \;:\; \omega \mapsto \Pi\, \omega \, .
\end{equation}
We decompose $\omega$ into
\be
\omega= \omega_0+ \omega'\, ,
\ee
where
$\omega_0$ is harmonic and $\omega'$ is in $\overline{ \mathrm{Harm}^r(\mathcal{M}) }$.
Now acting with $\triangle$
gives
\be
\triangle \omega= \triangle \omega '=  \triangle ' \omega '\,,
\ee
 and as $ \triangle '$ is invertible on $\overline{ \mathrm{Harm}^r(\mathcal{M}) }$
 we have
 \be
 \omega '= ( \triangle ' )^{-1} \triangle \omega\, .
 \ee
 But this is also $(\one-\Pi )  \omega$,
 so we conclude that the projector $\Pi$ is 
 \be\label{eq:pirforms}
 \Pi= \one -  ( \triangle ' )^{-1} \triangle\, .
 \ee
As the image of $\triangle$ is in $\overline{ \mathrm{Harm}^r(\mathcal{M}) }$, this is well-defined. Lastly, since $\omega' \in \overline{ \mathrm{Harm}^r(\mathcal{M}) }$ we can write
\begin{equation}
    \omega' = \triangle ( \triangle ' )^{-1} \omega' = d \big( d^\dagger ( \triangle ' )^{-1} \omega' \big) + d^\dagger \big( d ( \triangle ' )^{-1} \omega' \big)
\end{equation}
and this finishes the Hodge decomposition $\omega = d \Lambda + d^\dagger \beta + \omega_0$ of $\omega$: using the projector $\Pi$ defined in \eqref{eq:pirforms}, on can take $\omega_0 = \Pi \omega$, $\Lambda = d^\dagger ( \triangle ' )^{-1} (\one-\Pi) \omega$ and $\beta = d ( \triangle ' )^{-1} (\one-\Pi) \omega$.

\paragraph{A field theory model.} We now turn to $r$-form gauge fields on $\cM$. We introduce an  $r+1$ form field strength $F$ which is closed,
\be
dF= 0\, .
\ee
The  free action is given by
\be
S_0=\frac 1 2 (F,F)
\ee
Then by the Hodge decomposition theorem $F$ has a unique decomposition as
\be
F=dA + F_0
\ee
where $A$ is   a globally defined $r$-form and $F_0 $ is a harmonic $r+1$ form. $F_0$ determines the cohomology class $[F]$ of $F$ and then the theory is parameterised by
the
globally defined $r$-form
$A$.
The action is now
\be
S_0=\frac 1 2 (dA,dA)+\frac 1 2  (F_0,F_0)
\ee
and $(F_0,F_0)
$ is a constant depending only on the cohomology class $[F]$.
There is a gauge symmetry
\be
\delta A=d\Lambda
\ee
and the
field equation for $A$ is
\be 
d^\dagger d A=0\, .
\ee
The gauge symmetry is $(r-1)$-stage reducible, so that there are ``gauge symmetries for gauge symmetries''. Indeed, the variation $\delta \Lambda = d \lambda$ leaves $\delta A$ invariant, and the variation $\delta \lambda = d \rho$ leaves $\delta \Lambda$ invariant, and so on. The form degree of the gauge-for-gauge parameters is reduced by one at each stage, so that this chain eventually terminates with a scalar parameter (see diagram \eqref{chaimcomplex} below). This implies some subtleties upon gauge-fixing which we will discuss briefly below.

Now, since $A$ is a globally defined form, we can use the Hodge decomposition theorem and write
\be \label{eq:decompA}
A= d \Lambda + d^\dagger \beta + A_0
\ee
for some $\Lambda \in \Omega^{r-1}(\cM)$, $\beta \in \Omega^{r+1}(\cM)$ and $A_0 \in \mathrm{Harm}^r(\cM)$. 
Then the field equation implies that  $d^\dagger \beta =0$: indeed, it reduces to
\begin{align}
    d^\dagger d A = d^\dagger d d^\dagger \beta = \triangle (d^\dagger \beta) = 0
\end{align}
since $d\Lambda$ and $A_0$ are closed and $d^\dagger \beta$ is co-closed. This says that $d^\dagger \beta$ is harmonic, and therefore it must be zero (in the decomposition \eqref{eq:decompA}, the harmonic part of $A$ is only contained in $A_0$). Therefore, the general solution of the field equation is 
\be
A= d \Lambda   + A_0\;\ \qquad \triangle A_0 =0\;. 
\ee
Thus the solutions are harmonic forms modulo gauge transformations and the space of solutions is then the space of cohomology classes of harmonic forms.
However, each cohomology class has a unique harmonic representative so that the space of solutions is precisely the de Rham cohomology space $H^r ({\mathcal{M}} )$.

We can now enlarge this model by introducing further fields, denoted generically by $\phi$,
with the action
\be\label{eq:actionJA0}
S= \frac 1 2 (dA,dA)+ S_\phi + (A,J)
\ee
where $S_\phi $ is any action for the fields $\phi$ alone and $J(\phi )$ is some $r$-form current constructed from the fields $\phi$ which is required to be co-closed,
\be d^\dagger J(\phi) =0 \, ,\ee
to ensure the gauge-invariance of the last term. We also have absorbed the constant $(F_0,F_0)$ into $S_\phi$. We now wish to integrate out the field $A$, in order to obtain an effective theory for the fields $\phi$ alone. 

Let us first decompose $A$ as
\be A=A_0+A'
\ee
where
\be
A_0= \Pi A \in \mathrm{Harm}^r(\mathcal{M})
\ee
is the harmonic part of $A$ and 
\be A' =(\one- \Pi )A \in\overline{ \mathrm{Harm}^r(\cM)}\, .
\ee
The action then becomes
\be\label{eq:actionJA}
S=\frac 1 2 (dA',dA')+ S_\phi + (A',J')+ (A_0,J_0)
\ee
where we have used the orthogonality of the Hodge decomposition, the fact that $A_0$ is closed since it is harmonic, and defined
\be
J_0= \Pi J , \qquad J' =(\one- \Pi )J\, .
\ee

Note that the field $A'$ has a kinetic term, while the zero-mode part $A_0$ only appears as a Lagrange multiplier imposing the constraint $J_0(\phi) = 0$ on the fields $\phi$. The equation of motion for $A'$ reads
\be d^\dagger d A' + J' = 0 \, ,\ee
with general solution
\begin{equation}\label{generalA'}
    A' = - (\triangle')^{-1} J' + d \Lambda\, .
\end{equation}
So, $A'$ is determined in terms of $J'$ only up to a gauge transformation, as expected: because of the gauge symmetry, the kinetic operator $d^\dagger d$ appearing here is not invertible. 
{Reinserting the general solution (\ref{generalA'}) into the action, the $\Lambda$ contributions drop out as a consequence of gauge invariance, leaving 
\be\label{eq:Seff0}
\widehat{S} =  S_\phi - \frac 1 2(J',(\triangle ')^{-1}J')+ (A_0,J_0)
\ee
where one must use $d^\dagger J' = 0$ and the fact that $(\triangle ')^{-1}$ commutes\footnote{A formal proof proceeds as follows. From the definition of the Laplacian (\ref{eq:lapdef}) it immediately follows that 
$d\triangle=\triangle d${, and therefore also $d\triangle'=\triangle' d$ using the Hodge decomposition theorem}. Acting on $\triangle' (\triangle')^{-1}=1$ with $d$ from the left and the right we have $d\triangle' (\triangle')^{-1}= \triangle' (\triangle')^{-1}d$ and so 
 \be
  \triangle' d (\triangle')^{-1} = \triangle' (\triangle')^{-1}d  \;, 
 \ee
which implies $d(\triangle')^{-1}=(\triangle')^{-1}d$. {The same holds for $d^\dagger$.}} with $d$ and $d^\dagger$. In the following we will briefly explain how to obtain this, first, via homotopy transfer and, second, via 
a careful BV-BRST analysis using gauge fixing.}

\paragraph{Homotopy transfer.} 
We will now realize the integrating out of $A'$ as homotopy transfer. The underlying chain complex for the $r$-form gauge fields  reads 
   \be\label{chaimcomplex}
\begin{array}{ccccccccccc} 
 0 \,\longrightarrow\,
  \Omega^{0} \, \xlongrightarrow{d}\, \Omega^1 \,\xlongrightarrow{d}\,   \cdots  \,\xlongrightarrow{d}\,
\Omega^{r-1} 
\, \xlongrightarrow{d}\, \Omega^{r}    \, \xlongrightarrow{d^{\dagger}d}\, \Omega^{r} 
\, \xlongrightarrow{d^{\dagger}} \,\Omega^{r-1} \,\xlongrightarrow{d^{\dagger}} \,\cdots \,\longrightarrow \,0\;, 
\end{array}
\ee
where, as above, $\Omega^p$ denotes the space of $p$-forms. 
Here, $\Omega^r$ is interpreted as the space of $r$-form gauge fields but also as the space of its $r$-form field equations. Furthermore, 
the $\Omega^{i}$ for $i\leq r-1$ are interpreted as the space for gauge parameters, 
gauge for gauge parameters, etc., on the left half of the diagram, while they are interpreted as the spaces of Noether identities, Noether for Noether identities, etc., 
on the right half of the diagram. Thus, in terms of our $L_{\infty}$ conventions, the chain complex $(X,\partial)$ is defined by 
 \be
 \begin{split}
  i \geq 1\,: \quad X_i \ &= \ \Omega^{r-i}\;, \quad \quad \, \partial \ = \ d \,, \\
  i=0\,: \quad X_0 \ &= \ \Omega^{r}\;, \quad\qquad  \partial \ = \ d^{\dagger}d\, , \\
  i\leq -1\,: \quad X_i  \ &= \ \Omega^{r+i+1}\;, \quad\, \partial \ = \ d^{\dagger}\,.
 \end{split}
 \ee
Note, in particular, this is not the usual de Rham complex, as the differentials are only on the left half of the diagram given by 
the de Rham differential. Nevertheless, the above differential obeys $\partial_i\circ \partial_{i+1}=0$ for all $i$, as a consequence 
of $d^2=0$ and $(d^{\dagger})^2=0$. 

In order to formulate homotopy transfer we first define the projector $p: X\rightarrow \bar{X}$ for an arbitrary form $\omega$ in the above complex as 
 \be\label{harmonicprojector}
  p(\omega) \equiv \Pi(\omega) = \omega_0\;, 
 \ee
where, as above, $\omega_0$ denotes the harmonic part. The inclusion $\iota:\bar{X}\rightarrow X$ is trivial, just viewing a harmonic form as an element of the space 
of general forms. We then have $p\iota=\one_{\bar{X}}$. On the other hand, 
 \be\label{firststepinHomotopy}
   (\iota p -\one)(\omega)=-(\one-\Pi)\omega= -\omega'\,, 
 \ee   
   and we now have 
to define homotopy maps $h_i: X_i \rightarrow X_{i+1}$, so that the homotopy relations are obeyed. 
These maps are given by 
  \be\label{harmonicHomotopy}
 \begin{split}
  h_i  \ &= \ -(\triangle')^{-1} d^{\dagger}\qquad \qquad \text{for}\quad i=0,1,2, \ldots\;, \\
  h_{-1} \ &= \ -(\triangle')^{-1}(\one-\Pi)\;, \\
  h_i \ &= \ -(\triangle')^{-1} d\qquad\qquad\;\;  \text{for}\quad i=-2,-3,-4, \ldots\;. 
 \end{split} 
 \ee
Note that $(\triangle')^{-1}$ is always well-defined, since both $d$ and $d^{\dagger}$ map into the space of non-zero modes  and since for 
$h_{-1}$ the projector $(\one-\Pi)$ onto the space of non-zero modes was inserted. 
We now verify the homotopy relation acting on $\omega\in X_i$ for $i\geq 1$, in which case $\partial= d$, $h=-(\triangle')^{-1}d^{\dagger}$: 
 \be
 \begin{split}
  \partial(h \omega)+h(\partial\omega ) &= d(-(\triangle')^{-1} d^{\dagger}\omega) -(\triangle')^{-1}d^{\dagger}d\omega\\
  &=  -(\triangle')^{-1}(dd^{\dagger} + d^{\dagger}d)\omega\\
  &= - (\triangle')^{-1}\triangle \omega\\
  &=  - (\triangle')^{-1}\triangle' \omega'\\
  &= -\omega'\;, 
 \end{split} 
 \ee 
where we used that $d$ and $(\triangle')^{-1}$ commute. This agrees with (\ref{firststepinHomotopy}) and so we have verified the homotopy relation on $X_i$ for $i\geq 1$. 
For $\omega \in X_{i}$ with $i\leq -2$, in which case  $\partial = d^{\dagger}$ and $h=-(\triangle')^{-1}d$, the proof is precisely analogous. 
Special care is required for the borderline cases $X_0$ and $X_{-1}$. For $\omega\in X_{0}$ we have 
 \be
 \begin{split}
  \partial_1(h_0\omega) + h_{-1}(\partial_0\omega) &= d(-(\triangle')^{-1}d^{\dagger}\omega) - (\triangle')^{-1}(\one-\Pi)d^{\dagger}d\omega\\
  &=  -(\triangle')^{-1}\triangle' \omega' \\
  &=-\omega' \;, 
 \end{split}
 \ee
where we used that $(\one-\Pi)$, which projects onto the non-zero modes, acts as the identity on $d^{\dagger}d\omega$.\footnote{{ This follows from the Hodge decomposition theorem: the form $d^\dagger (d \omega)$ is coexact and therefore has no harmonic component, so $\Pi d^\dagger d \omega = 0$.} } 
This establishes the homotopy relation on $X_0$. The proof of the homotopy relation on $X_{-1}$ is similarly straightforward, 
thereby completing the proof that the projector (\ref{harmonicprojector}) onto harmonic forms together with  (\ref{harmonicHomotopy}) 
defines a homotopy transfer.

Having shown that the projection to harmonic forms (zero modes) can be interpreted as homotopy transfer we now sketch 
how to obtain from this the effective action (\ref{eq:Seff0}) obtained after integrating out $A'$. 
We first note that the chain complex has to be extended by the extra fields $\phi$, together with their gauge parameters, etc., if present. 
We denote the fields collectively by ${\cal A}=(A,\phi)$. Consider, for simplicity, the case where the current $J$ is bilinear in $\phi$. We then have a 2-bracket $[{\cal A}, {\cal A}]$ on fields whose projection onto the $r$-form part is given by 
 $-2J$, so as to match the dictionary (\ref{MaurerCartanEOM}) 
 and to yield the cubic couplings $(A,J)$ in the action. 
 The projection on the full field space is given by $p({\cal A})=\overbar{\cal A}=(A_0,\phi)$ (together 
 with a suitable extension to the full chain complex). 
 We recall that the homotopy transported 2-bracket is given by
 \be
  [\overbar{\cal A}, \overbar{\cal A}]=p[\iota(\overbar{\cal A}), \iota(\overbar{\cal A})] \;, 
 \ee 
where $p$ projects onto the harmonic part. This implies that the current is projected onto the harmonic part, so that the cubic couplings reduce to $(J_0, A_0)$, as expected from (\ref{eq:Seff0}).
Similarly, there will be an induced quartic coupling encoded in the transported 3-bracket, in precise  analogy to the models in the 
previous section (see the discussion starting with (\ref{transportedBRacket})).

\paragraph{BV-BRST analysis.}
In order to integrate out $A'$ in {the path-integral formalism}, we must gauge-fix the gauge symmetry.
The most convenient for our purposes is to use the Gaussian gauge-fixing term $\frac{1}{2}(d^\dagger A, d^\dagger A)$, which gives the kinetic term
\begin{align}
    \frac{1}{2}(dA',dA') + \frac{1}{2}(d^\dagger A',d^\dagger A') &= \frac{1}{2} (A', d^\dagger d A') + \frac{1}{2}(A', d d^\dagger A') \\
    &= \frac{1}{2} (A', \triangle' A')
\end{align}
where $A_0$ again drops out since it is co-closed. {The} action \eqref{eq:actionJA} then becomes
\begin{equation}\label{eq:actionJAgf}
    S_\text{gf} = \frac 1 2 (A',\triangle' A') + S_\phi + (A',J') + (A_0,J_0) + S_{\text{ghost}}\, ,
\end{equation}
where $S_{\text{ghost}}$ is the action for the whole spectrum of ghosts (including extraghosts, ghosts-for-ghosts etc.) necessary to properly implement this gauge-fixing procedure in the case of a reducible theory. For example, in the case $r=2$ one has two one-form ghosts $C$, $\overbar{C}$ of degree $1$ and $-1$, and three scalar (0-form) ghosts $c$, $\bar{c}$, $\eta$ of degree $2$, $-2$ and $0$ (see e.g.~\cite{Batalin:1984jr} or \cite{Henneaux:1992ig} where this example is treated in detail). The ghost action then simply reads
\begin{equation}
    S^{(r=2)}_{\text{ghost}} = \frac{1}{2} (\overbar{C},\triangle' C) + \frac{1}{2} (\bar{c},\triangle' c) + \frac{1}{2} (\eta, \triangle' \eta)
\end{equation}
with invertible kinetic operators for all the fields. In these models, they are decoupled from all the other fields (this would not be the case if the gauge symmetry were non-abelian or would only close off-shell, for example).

Now, the equation of motion for $A'$ simply reads $\triangle' A' + J' = 0$, with unique solution $A' = - (\triangle')^{-1} J'$. Plugging this solution back into the action then gives the effective action
\be\label{eq:Seff}
\widehat{S} =  S_\phi - \frac 1 2(J',(\triangle ')^{-1}J')+ (A_0,J_0)
\ee
where we also integrated out the corresponding ghost sector, which in this case simply corresponds to setting them all to zero.

{This coincides with the result \eqref{eq:Seff0} obtained previously, as it should. This action still contains the zero-modes $A_0$. As mentioned above, they appear as a Lagrange multiplier imposing the constraint
\be J_0(\phi) = 0 \, . \ee
Of course, if the specific form of $J$ is such that this is identically satisfied, then the last term of \eqref{eq:Seff} is absent and $A_0$ simply drops out from the action.}

\paragraph{Truncating zero-modes and homotopy transfer.}  Let us now make the link with the notation of section \ref{sec:zeromodesgeneral}.

{When attempting to integrate out $A$ in its entirety from action \eqref{eq:actionJA0}, or its gauge-fixed version \eqref{eq:actionJAgf}, the projector $P$ acts as
\begin{equation}
    P A = 0\, , \quad P \phi = \phi\, , \quad P\cC = 0\, ,
\end{equation}
where $\cC$ denotes any of the ghost fields. Then, the projector $\widehat{P} = P + (\one-P)\Pi$ acts as
\begin{equation}
    \widehat{P} A = A_0\, , \quad \widehat{P} \phi = \phi\, , \quad \widehat{P}\cC = \cC_0\, ,
\end{equation}
i.e.~projects out $A'$, $\cC'$ but \emph{keeps} the zero-modes of the sector to be integrated out. As shown above (where indeed $\iota p = \widehat{P}$), homotopy transfer then produces the $L_\infty$ algebra of the action $\widehat{S}$ {of \eqref{eq:Seff0}}, where $A_0$ still appears since the homology did not change. 

Now, if $J_0 = 0$ identically, the zero-modes $A_0$ drop out from the action and they can be consistently truncated out in the sense of section \ref{sec:zeromodesgeneral}, leading to a compatible $L_\infty$-algebra structure on $PX$ with morphism $E$ given by the natural inclusion. If $J_0 \neq 0$, however, setting $A_0 = 0$ is inconsistent. This illustrates the fact that, while homotopy transfer to $\widehat{P}X$ is always possible, the consistency of the truncation of zero-modes depends on the precise theory at hand.

Lastly, the projector $\widetilde{P} = \one - \Pi(\one-P)$ of section \ref{sec:zeromodesgeneral} acts as
\begin{equation}
    \widetilde{P} A = A'\, , \quad \widetilde{P} \phi = \phi\, , \quad \widetilde{P}\cC = \cC'\, ,
\end{equation}
i.e.~it only projects out the zero-modes $A_0$, $\cC_0$ of the sector to be integrated out. Again, when $J_0=0$ there is an $L_\infty$-algebra structure on $\widetilde{P}X$ which is a consistent truncation, corresponding to the action \eqref{eq:actionJA} without the last term. Homotopy transfer can then be used to integrate out $A'$, $\cC'$ and produce the $L_\infty$-structure on $PX$.
}

\section{Closed strings on a torus}
\label{sec:stringstorus}

In this section we briefly review some general facts  about string theory in toroidal backgrounds, establishing   our notation. 
We follow  the textbook \cite{Blumenhagen:2013fgp}; our formulas agree with those given there if we set $\hbar=1\,,\alpha'=2$. Our conventions  are also compatible with those of the paper \cite{Zwiebach:1992ie}.

Consider the  closed bosonic string propagating in a target space metric that is the direct product of  $n$-dimensional Minkowski space with a $d$-dimensional torus,
 with constant target space  metric $G_{ij}$ and vanishing $B$-field. 
The classical worldsheet action with worldsheet metric $\gamma_{\alpha\beta}$ and coordinates $\sigma^{\alpha}=(\tau,\sigma)$ then reads 
 \be
  S=-\frac{1}{8\pi}\int_{0}^{2\pi}d\sigma \int d\tau \sqrt{\gamma}\gamma^{\alpha\beta}\partial_{\alpha}X^i\partial_{\beta}X^{j} G_{ij}\,,  
 \ee
where $\sigma$ has period $2\pi$, 
\be
\sigma\sim \sigma + 2\pi\,.
\ee
We then split indices and the embedding scalars $X:\Sigma_2\to T^d\times \mathbb R^n$ according to 
\be
X^i=\{X^a,X^\mu\}\,,\qquad i=0,1,\dots 25\,,\quad a=1,2,\dots d\,,
\ee
where $X^\mu$ are the Minkowski coordinates and $X^a$ are the   toroidal coordinates, subject to the periodicity 
\be
X^a\sim X^a+2\pi\,.
\ee
Note that the $X^i$ are dimensionless in our conventions. 
The $(d+n)$-dimensional spacetime metric $G$, which encodes in particular all information of the torus geometry including physical lengths, 
 has the block diagonal decomposition 
\be
G_{ij}=\begin{pmatrix}\eta_{\mu\nu} & 0 \\0  & G_{ab}\end{pmatrix}\,. 
\ee
Here $\eta_{\mu\nu}$ is the Minkowski metric and $G_{ab}$ is the torus metric 
(where, in our convention, the rectangular torus at the self-dual radius corresponds to $G_{ab}=2\delta_{ab}$).

We now turn to the quantisation of this theory, assuming a Euclidean worldsheet metric, and using 
radial quantisation with the usual holomorphic coordinate
\be
z=\exp(\tau-i\sigma)\,,
\ee
where $\tau$ is Euclidean time.\footnote{As is usually done, we will formally regard $z, \bar z$ as independent complex variables, but refer to dependence on $z$ as holomorphic
and dependence on $\bar z$ as anti-holomorphic.}
The embedding scalars  $X^i=X^i(z,\bar z)$ decompose into holomorphic and anti-holomorphic parts, 
\be
X^i(z,\bar z)\equiv X^i(z)+\bar X^i(\bar z)\,,
\ee
which may be expanded into modes as follows 
\be
\begin{split}
X^i(z)&\equiv x^i -i\log(z)\alpha_0^i + i\sum_{n\neq 0}\frac{1}{n}\frac{\alpha_n^i}{z^n}\,,\\
\bar X^i(\bar z)&\equiv \bar x^i -i\log(\bar z)\bar\alpha_0^i + i\sum_{n\neq 0}\frac{1}{n}\frac{\bar\alpha_n^i}{\bar z^n}\,.
\end{split}
\ee
As usual, the operator
\be
i\pd X^i(z)=\sum_n z^{-1}\frac{\alpha_n^i}{z^n}
\ee
is a conformal primary of conformal weights $h=1,\bar h=0$. 
The oscillators $\alpha^i_n$ satisfy the commutation relations
\be
\label{matteroscillatorcommut}
[\alpha^i_m,\alpha^j_n]=m G^{ij}\delta_{m+n}\,, \qquad m,n\in\mathbb Z\,, 
\ee 
and the antiholomorphic oscillators $\bar \alpha^i_n$ satisfy   similar relations. 

We choose conformal gauge and introduce the usual 
$bc$ and $\bar b\bar c$ ghost systems. For the holomorphic ghosts we have the mode expansions
\be
b(z)=\sum_{n}\frac{1}{z^{n+2}}b_n\,,\quad c(z)= \sum_{n}\frac{1}{z^{n-1}}c_n\,, 
\ee
for the conformal primaries $b(z)$ and $c(z)$ of weights ($h=2$, $\bar h=0$) and ($h=-1$, $\bar h=0$), respectively. The oscillators $b_n$, $c_n$ satisfy the commutation relations
\be
\label{ghostosscillatorcommut}
\{b_m,c_n\}=\delta_{m+n}\,.
\ee

The oscillator commutators \eqref{matteroscillatorcommut} and \eqref{ghostosscillatorcommut} are the same for a non-compactified background and a toroidal background. The difference lies in the zero modes of the matter oscillators. If all dimensions were   non-compact,  we would 
 have $\alpha_0^i=\bar\alpha_0^i\propto G^{ij}\mathrm p_j$, where $\mathrm p_i$ is the string momentum. For our toroidal background, the presence of winding modes 
 make the holomorphic and antiholomorphic zero-modes for the torus  independent:
\be
\label{matterzeromodesintermsofwindingandmomentum:appendix}
\bar \alpha_0^a- \alpha_0^a= {\mathrm w}^a\,,\qquad \alpha_0^a+\bar\alpha_0^a=2 {\mathrm p}_b G^{ab}\,,\qquad \alpha_0^\mu=\bar\alpha_0^\mu= \eta^{\mu\nu}\mathrm p_\nu\,.
\ee
The  torus momentum and winding operators
\be
\mathrm p_a=\tfrac{1}{2} G_{ab}(\alpha_0^a+\bar\alpha_0^a),\qquad \mathrm w^a=\bar\alpha^a-\alpha^a\,, 
\ee
are independent and have integer eigenvalues in our conventions. The torus momentum is the canonical conjugate to the string centre-of-mass position \be
\label{centreofmassposition}
\mathrm x^a=x^a+\bar x^a\,,
\ee and $\mathrm w^a$ is conjugate to the dual coordinate 
\be
\label{dualposition}
\tilde {\mathrm x}_a=\tfrac{1}{2}G_{ab}(\bar x^b-x^b)\,.
\ee
The factors of $2$ in these formulae reflect the fact that the self-dual radius is at $G_{aa}=2$ in our convention with $\alpha'=2$.
It will also be convenient to express these in terms of independent (anti)holomorphic zero-modes
\be
\label{leftrightmomenta}
p_a=G_{ab}\alpha^b_0\,,\qquad \bar p_a=G_{ab}\bar \alpha^b_0\,,\qquad \mathrm{p}_a = \tfrac{1}{2}(p_a + \bar{p}_a)\, ,
\ee
with
\be
\label{leftrightpositionmomentumcommutator:appendix}
[x^a,p_b]=i\delta^a_b\,,\qquad [\bar x^a,\bar p_b]=i\delta^a_b\,.
\ee

It is sometimes interesting to consider an alternative quantisation procedure in which
the toroidal position zero-modes are taken  to be non-commutative with
\be
\label{noncommutativepositionzeromodes}
[x^a,\bar x^b]=i c \,G^{ab}\,,\qquad c\in\mathbb R\,,
\ee
where  $c$ is a new parameter, so that the familiar commutative case is recovered in the limit $c\rightarrow 0$; see \cite{Freidel:2017wst} for a discussion and \cite{Tseytlin:1990nb,Lizzi:1997xe} for related ideas.
The formulas \eqref{matteroscillatorcommut}, \eqref{ghostosscillatorcommut}, \eqref{leftrightpositionmomentumcommutator:appendix}, and \eqref{noncommutativepositionzeromodes} collect the non-vanishing commutators for this system. Here we will use the usual commutative ($c=0$) quantisation but we will comment on the alternative quantisation where relevant.

Next, we collect formulas for 
the holomorphic (antiholomorphic) Virasoro generators $L_n$ ($\bar L_n$) and the BRST charge $Q$: 
\begin{align}
\label{virasorosTorus}
\begin{split}
L_n&\equiv L_n^{\rm matter}+L_n^{\rm ghost}\;,\\
L_n^{\rm matter}&\equiv\sum_m \frac{1}{2} {:}\; \alpha^i_{n-m}G_{ij}\alpha^j_{m}\;{:}\;,\qquad
L_n^{\rm ghost}\equiv\sum_m(n+m) {:}\; b_{n-m}c_m \;{:}\,,\\
Q_{\rm B}& \equiv\sum_n {:}\; c_{-n}(L_n^{\rm matter}+\tfrac{1}{2}L_n^{\rm ghost})\;{:} + {\rm antiholomorphic.}
\end{split}
\end{align}
Here $:\;\; :$ denotes conformal normal ordering (see \cite[section 4.2]{Blumenhagen:2013fgp} for details), so that e.g.
\be
L_0^{\rm ghost}=-1 + \sum_{n>0}n(b_{-n}c_n + c_{-n}b_n)\,.
\ee
The number operators $N$ and $\bar N$ are defined as
\be
\label{levelnumberexplicit}
\begin{split}
N&=\sum_{n>0}\alpha_{-n}^i G_{ij}\alpha_n^j + n(b_{-n}c_n+c_{-n}b_n)\,,\\
 \bar N&=\sum_{n>0}\bar \alpha_{-n}^i G_{ij}\bar\alpha_n^j + n(\bar b_{-n}\bar c_n+\bar c_{-n}\bar b_n)\,.
\end{split}
\ee
In terms of these, $L_0$ and $\bar L_0$ can be written as
\be
\label{leveloperators}
L_0=N-1 +\frac{1}{2}\alpha_0^i G_{ij}\alpha_0^j\,,\qquad \bar L_0=\bar{N}-1 +\frac{1}{2}\bar\alpha_0^i G_{ij}\bar\alpha_0^j\,.
\ee
Defining  $L_0^+=L_0+\bar L_0$ and $L_0^-=L_0-\bar L_0$ one may express the string mass-shell condition as $L_0^+=0$ and the level-matching constraint as $L_0^-=0$. 
Written in terms of momenta and winding we have 
\be
\label{stringL0pmexpressions}
\begin{split}
L_0^+&=(N+\bar N -2)+\mathrm p_\mu \mathrm p_\nu\eta^{\mu\nu}+ \mathrm p_a \mathrm p_b G^{ab}+\tfrac{1}{4}\mathrm w^a \mathrm w^b G_{ab}\,,\\
 L_0^-&=(N-\bar N)-\mathrm p_a \mathrm w^a\,.
\end{split} 
\ee
Note that only compact momenta $\mathrm p_a$ and windings $\mathrm w^a$ appear in the level-matching condition, while the full set of momenta and windings appear in the mass-shell condition. (The numerical factors in $L_0^+$ correspond to the self-dual radius being $G_{aa}=2$ in our conventions.) The expression for $L_0^-$ does not depend on the torus metric, while $L_0^+$ does.

The $SL(2;\mathbb C)$-invariant vacuum  state $|{\bf 0}\rangle$  is annihilated by $L_0,L_{\pm1}$ and $\bar L_0,\bar L_{\pm1}$.
It can be regarded as the vacuum in the asymptotic past (at $z = 0$), with no operator insertions.
 The $SL(2;\mathbb C)$-invariant vacuum $|{\bf 0}\rangle$ is annihilated by   oscillators  with  high enough mode number:
 \be
\label{sl2cannihilationconditions}
\alpha_n^i|{\bf 0}\rangle=b_{n-1}|{\bf 0}\rangle=c_{n+2}|{\bf 0}\rangle=0\qquad\forall n\geq 0\,,
\ee 
as well as the analogous conditions involving barred oscillators.
The standard physical vacuum is then
\be
|\!\downarrow\downarrow\rangle=c_1\bar c_1 |{\bf 0}\rangle\,.
\ee
If  (as in \cite{Zwiebach:1992ie})
 $|{\bf 0}\rangle$ is taken to have ghost number zero, then $|\!\downarrow\downarrow\rangle$ has ghost number 2.
 States with momentum $\mathrm p_i = (\mathrm p_\mu, \mathrm p_a)$ and winding number $\mathrm{w}^a$
 ($\mathrm p_\mu \in\mathbb R^n\,,\mathrm p_a,\mathrm w^a\in\mathbb Z^{2d}$)
 can be obtained by acting with
\be
 \exp(ik_i \hat x^i)\exp(i\bar k_i\hat{\bar x}^i)
 \ee
where
\be
\label{kbarkintermsofintegers}
k_a=\mathrm p_a-\tfrac{1}{2} \mathrm w^b G_{ab}\,,\quad \bar k_a=\mathrm p_a+\tfrac{1}{2}\mathrm w^b G_{ab}\,,\qquad k_\mu=\bar k_\mu=\tfrac{1}{2}\mathrm p_\mu\, .
\ee
For example
\be
\label{nottachyons}
|{\bf 0}; \mathrm p,\mathrm w\rangle\equiv \exp(ik_i \hat x^i)\exp(i\bar k_i\hat{\bar x}^i)|{\bf 0}\rangle\,, 
\ee
while
\be
|\!\downarrow\downarrow; \mathrm p,\mathrm w\rangle\equiv \exp(ik_i \hat x^i)\exp(i\bar k_i\hat{\bar x}^i)|\!\downarrow\downarrow\rangle\,
\ee
is a tachyon state with momentum $\mathrm{p}_i$ and winding $\mathrm{w}^a$. Other string states are then obtained by acting on this state with oscillators. The ``doubly massless'' states $\ket{\Psi}$ at level $N=\bar N=1$ (with $b_0-\bar b_0=0$) will be especially important for us. They are constructed via oscillators as
\be
\label{doublymassless}
\begin{split}
\ket{\Psi} =  \int [d\mathrm{p}d\mathrm{w}] ~ \Bigl(  & - {\frac{1}{2}} 
e_{ij}  \,\alpha_{-1}^i \bar \alpha_{-1}^j \, c_1 \bar c_1
 + e  \, c_1 c_{-1}    +  \bar e   
\, \bar c_1 \bar c_{-1} \\[0.5ex]
&+ i f_i  \, c_0^+ c_1 \alpha_{-1}^i 
+ i \bar f_j  \, c_0^+ \bar c_1 \bar \alpha_{-1}^j\Bigr)\, |{\bf 0}; \mathrm p,\mathrm w\rangle \, ,
\end{split}
\ee
where the coefficients $\{e_{ij}, e_i,\bar e_i, f_i,\bar f_i\}$ all depend on the momenta and windings $\mathrm{p,w}$, and $\int [d\mathrm{p}d\mathrm{w}]$ is an integral over the $n$ non-compact momenta and the $2d$ compact momenta and windings.

The dual  $SL(2;\mathbb C)$-invariant vacuum  $\langle {\bf 0}|$ is a state in the dual space representing the vacuum in the asymptotic future ($z=\infty$)
satisfying
\be
\langle {\bf 0}|\alpha_{-n}^i=\langle {\bf 0}|b_{1-n}=\langle {\bf 0}|c_{-2-n}=0\qquad\forall n\geq 0\,.
\ee
The duality pairing is  
\be
\langle{\bf 0}| c_{-1}\bar c_{-1} c_0^+c_0^- c_1\bar c_1|{\bf 0}\rangle=1
\ee
Introducing momentum and winding, we then have the basic overlap  
\be
\label{overlap}
\langle{\bf 0};\mathrm p',\mathrm w'| c_{-1}\bar c_{-1} c_0^+c_0^- c_1\bar c_1|{\bf 0};\mathrm p,\mathrm w\rangle\equiv(2\pi)^{n+2d}\delta^n(\mathrm p_\mu-\mathrm p'_\mu)\delta^d(\mathrm p_a-\mathrm p'_a)\delta^d(\mathrm w^a-\mathrm w'^a)
\ee
that agrees with reference \cite{Hull:2009mi} (up to a sign).

\section{Closed string fields on a torus}
\label{sec:sft}

We set up non-polynomial covariant closed string field theory \cite{Kugo:1989tk,Zwiebach:1992ie}  for the torus CFT in order to address the construction of Double Field Theory later. Homotopy transfer only requires explicit knowledge of the quadratic genus-zero structure, which makes this discussion relatively straightforward. However, we will encounter some new subtleties around the reflector state on toroidal backgrounds, closely related to the (non)commutativity of vertex operators. The reflector state appears from the \lf-algebra point of view through the cyclic inner product.

We begin with general features of the construction of the \lf-algebra. The basic result of tree-level covariant bosonic closed string field theory \cite{Zwiebach:1992ie} is that
\begin{proposition} Given a ``matter'' CFT of central charge $+26$ (along with the universal $bc,\bar b\bar c$ Virasoro ghost sector of charge $-26$), there exist (graded symmetric, multilinear) brackets $b_n: X^n\to X$
\be
b_1\equiv \pd,b_2,b_3,\dots
\ee
and a non-degenerate, complex-linear inner product $\kappa$
\be
\kappa: X\times X\to  \mathbb C
\ee
on the space $X$ of string states that satisfy the level matching and $b_0-\bar b_0=0$ constraints
\be
\label{levelmatchingandantighostcondition}
x\in X\implies (L_0-\bar L_0)x=(b_0-\bar b_0)x=0\,,
\ee
so that
\be
(X,\{b_n\},\kappa)
\ee
together define a cyclic \lf-algebra.
\end{proposition}
\noindent This \lf-algebra provides a Lagrangian whose tree-level scattering amplitudes are the genus-zero string amplitudes. (The generalisation including arbitrary genus brackets is treated in \cite{Zwiebach:1992ie}; the algebraic structure is that of a ``loop'' or ``quantum'' \lf-algebra.) For our purposes we do not need to display the brackets $b_{n\geq 2}$, since the conditions for the validity of homotopy transfer onto a sub-complex of states only involve the 1-bracket $\pd$ and inner product $\kappa$ explicitly.

The grading on $X$ that is respected by the brackets $b_n$ and inner product $\kappa$ in the usual way is essentially the same as the ghost number:
\be
\label{ghosttodeggrading}
{\rm gh}\; x= 2-\deg x\,,
\ee
where we employ the homological degree convention on the \lf-algebra side, so all brackets $b_n$ have $\deg=-1$ (as in Section 2 of the prequel \cite{Arvanitakis:2020rrk}).

The 1-bracket $\pd$ is the worldsheet BRST charge
$Q_{\rm B}$ given in \eqref{virasorosTorus} for the torus CFT:
\be
\partial\equiv Q_{\rm B}=c_0L_0+\bar c_0 \bar L_0+\dots
\ee
Since $\{b_0,Q_{\rm B}\}=L_0\,,\{\bar b_0,Q_{\rm B}\}=\bar L_0$ due to general properties of the ghost system, $Q_{\rm B}$ is well defined on the space $X$ of string states \eqref{levelmatchingandantighostcondition} with level-matching and $b_0-\bar b_0$ constraints.

The cyclic inner product $\kappa$ is defined via what is often called a \emph{reflector state} $R$ by
\be
\label{cyclicdef}
\kappa(x_1,x_2)\equiv R\left(x_1, \tfrac{1}{2}(c_0-\bar c_0)x_2\right)\,.
\ee
$R$ is a bilinear form $R:X\times X\to\mathbb C$ that is graded-symmetric
\be
R(x_1,x_2)=(-1)^{x_1x_2}R(x_2,x_1)
\ee
and of degree $+2$ (or equivalently of ghost number $-6$). With these degree assignments, the bilinear form
\be
(x_1,x_2)\to\kappa(\pd x_1,x_2)
\ee
is of degree zero. This defines the string field kinetic term. The cyclic inner product $\kappa$ and 1-bracket $\pd$ are mutually compatible if this bilinear form is graded symmetric:
\be
\kappa(\pd x_1,x_2)=(-1)^{x_1x_2}\kappa(\pd x_2,x_1)\,\,.
\ee
We refer to \cite{Zwiebach:1992ie} for a proof of this relation via the definition \eqref{reflector:def} of the reflector.

 The reflector state is defined for any CFT through the state-operator correspondence and conformal inversion\footnote{For $\varphi=0$ this matches the torus CFT reflector state of Kugo and Zwiebach \cite{Kugo:1992md}.}
\be
\label{inversion:maintext}
I(z)=-e^{i\varphi}z^{-1},\quad I(\bar z)=-e^{-i\varphi}\bar z^{-1}\,.
\ee
Here the phase $\varphi$ is an arbitrary constant. If the state $x_1$ is created by the operator $\mathcal O_1(z,\bar z)$ acting on the $SL(2;\mathbb C)$-invariant vacuum $|{\bf 0}\rangle$:
\be
x_1=\lim_{z,\bar z\to 0}\mathcal O_1(z,\bar z) |{\bf 0}\rangle\,,
\ee
and similarly for $x_2$, then
\be
\label{reflector:def}
R(x_1,x_2)\equiv \lim_{z,\bar z,w,\bar w\to 0}\langle {\bf 0}| (I\circ\mathcal O_1)(z,\bar z) \mathcal O_2(w,\bar w)|{\bf 0}\rangle\,,
\ee
where $I\circ$ is the action of conformal inversion \eqref{inversion:maintext} on operators, {which acts on the argument but also involves a transformation of the operator itself that depends on the conformal weight}. The reflector state therefore depends on an arbitrary choice of phase $\varphi$, but there are further ambiguities that we discuss shortly.

The adjoint $\mathcal O^T$ with respect to $R$ is known as the \emph{BPZ} conjugate \cite{Belavin:1984vu}. (This is complex-linear as the $^T$ notation suggests, since $R$ and $\kappa$ are linear in both arguments.) On states $x\in X$ we define
\be
\label{transposedef}
x^T\equiv R(x,\--)\in X^\star\,.
\ee
In the convention
\be
\label{transposeconvention}
R(\mathcal O x_1,x_2)=(-1)^{x_1\mathcal O} R(x_1,\mathcal O^Tx_2),
\ee
we find the rules
\be
\label{BPZtransposerules}
(\mathcal Ox)^T=(-1)^{x\mathcal O} x^T\mathcal O^T\,,\quad (\mathcal O_1\mathcal O_2)^T=(-1)^{\mathcal O_1\mathcal O_2}\mathcal O_2^T\mathcal O_1^T
\ee
which lead to the usual formula for the BPZ conjugate of a string of normal-ordered oscillators hitting the vacuum state: 
\be
(\lambda \alpha_{-n_1} b_{-n_2} c_{-n_3}\cdots |{\bf 0}\rangle)^T= \langle{\bf 0}| \lambda \alpha_{-n_1}^T b_{-n_2}^T c_{-n_3}^T\cdots\,,\quad \lambda \in\mathbb C\,.
\ee
Given the collection of primary operators in the CFT, whose transformation under inversion $I\circ$ is known by definition, one can calculate the BPZ conjugate of any state via the rules \eqref{BPZtransposerules} and therefore calculate the reflector state $R$. Specifying the BPZ conjugation $\mathcal O^T$ is equivalent to specifying the reflector state $R$.

Easy oscillator gymnastics, along with the overlap \eqref{overlap} and the anticommutator
\be
\{b_0^+,c_0^+\}=1\,,\qquad b_0^+=b_0+\bar b_0\,,\quad c_0^+=\frac{1}{2}(c_0+\bar c_0)\,,
\ee
suffice to prove the following general properties of the cyclic inner product $\kappa$:
\begin{proposition}
\begin{enumerate}
\item The subspaces $\im{b_0^+}$ and $\im{c_0^+}$ are each maximally isotropic for $\kappa$:
\be
\label{SFTisotropicsubspaces}
\kappa(b_0^+x,b_0^+y)=\kappa(c_0^+x,c_0^+y)=0\qquad \forall x,y\in X\,.
\ee
\item The Virasoro generator \eqref{virasorosTorus} $L_0$ and level operator $N$ \eqref{levelnumberexplicit} are BPZ self-conjugate, and
\be
\label{SFTkappaadjointL0level}
\kappa(L_0 x,y)=\kappa(x,L_0 y)\,, \quad \kappa(Nx,y)=\kappa(x,Ny)\qquad \forall x,y\in X\,.
\ee
\end{enumerate}
\end{proposition}

\subsection{Ambiguities in the reflector, cocycles, and non-commutativity }

The definition of the reflector state and of BPZ conjugation suffers from an ambiguity beyond the choice of phase $\varphi$ in \eqref{inversion:maintext}. For the torus CFT this leads to a sign ambiguity for the BPZ conjugates of certain states carrying both momentum and winding. The resolution will suggest a recipe for cocycle sign insertions in all string vertices.

This subtlety is closely related to the old issue of momentum- and winding-dependent sign factors in the string vertices, known as ``cocycle'' sign factors, which affect the form of the reflector and string vertices for the torus background relative to those for an uncompactified background. Direct calculations in covariant HIKKO string field theory \cite{Hata:1985zu,Hata:1985tt} showed the Jacobi identities and the symmetry of (at least) the binary brackets are violated unless cocycle signs are inserted in both the binary bracket (cubic vertex) and the reflector state \cite{Hata:1986mz,Kugo:1992md}. A recipe for the insertion of cocycle signs in the quartic vertex and beyond appears missing from the literature.

Definition \eqref{reflector:def} of $R(x_1,x_2)$ is ambiguous because the operators $\mathcal O_1(z,\bar z),\mathcal O_2(z,\bar z)$ that respectively create the states $x_1,x_2$ are not uniquely specified. For a torus background, but not e.g.~for a Minkowski one, there are indeed multiple choices of vertex operators, corresponding to different insertions of \emph{cocycle operators}. Those are related to the non-commutativity, for a torus background, of  the vertex operators  that create the states \eqref{nottachyons}, which we now review. The ``naive'' vertex operator 
\be
\label{naive:def:maintext}
\mathcal V^\text{naive}_{k,\bar k}(z,\bar z)\equiv\,:\exp(ik \hat X(z))\exp(i\bar k \hat {\bar X}(\bar z)):
\ee
creates the state \eqref{nottachyons}, which we display again here:
\be
\label{tachyonstate}
|{\bf 0};\mathrm{p},\mathrm{w}\rangle\equiv \exp(i k \hat x)\exp(i\bar k \hat{\bar x})|{\bf 0}\rangle
\ee
of momentum ${\mathrm p}_k=(k+\bar k)/2$ and winding $\mathrm{w}_k=\bar k-k$. $\mathcal V^\text{naive}_{k,\bar k}$ has the following equal-time commutation relation with another such operator $\mathcal V^\text{naive}_{\ell,\bar \ell}(w,\bar w)$:
\be
\label{badcommutation}
\mathcal V^\text{naive}_{k,\bar k}(z,\bar z) \mathcal V^\text{naive}_{\ell,\bar \ell}(w,\bar w)|_{|z|=|w|+\varepsilon}=(-1)^{(\bar k\bar\ell-k\ell)} \mathcal V^\text{naive}_{\ell,\bar \ell}(w,\bar w) \mathcal V^\text{naive}_{k,\bar k}(z,\bar z)|_{|z|=|w|-\varepsilon}\,,
\ee
where $\bar k\bar\ell-k\ell={\mathrm p}_k {\mathrm w}_\ell+{\mathrm p}_\ell {\mathrm w}_k\in\mathbb Z$ is the standard $O(d,d)$ quadratic form on the lattice of momenta and windings. (We assumed here as is conventional that the position zero-modes commute.)
Therefore, even though  these states are bosonic, the operators creating them do not obey bosonic commutation relations.

This is an old issue and its resolution has been known for a long time \cite{Frenkel:1980rn,Goddard:1983at,Gross:1985rr,Sakamoto:1989ig,Erler:1991an,Sakamoto:1992ur,Horiguchi:1992sn,Sakamoto:1993bc,Sakamoto:1994nx,Landi:1998ii,Hellerman:2006tx}.{\footnote{We especially profited from the recent discussion of Freidel, Leigh and Minic \cite{Freidel:2017wst}, and from unpublished notes of Barton Zwiebach.}} The cure is to dress the naive  vertex operator with a cocycle operator $C_{k,\bar k}$:
\be
\label{correctedvertex:defmaintext}
\mathcal V_{k,\bar k}(z,\bar z)\equiv \mathcal V^\text{naive}_{k,\bar k}(z,\bar z)C_{k,\bar k}\,.
\ee
We will see shortly that $C_{k,\bar k}$ can be chosen to satisfy
\be
C_{k,\bar k}|\mathbf{0}\rangle=|\mathbf{0}\rangle\,,\quad \langle\mathbf{0}|C_{k,\bar k}=\langle\mathbf{0}|\,,
\ee
so $\mathcal V^\text{naive}_{k,\bar k}$ and $\mathcal V_{k,\bar k}$ create the same ket state \eqref{tachyonstate}. However,
\be
\lim_{z,\bar z\to 0}\langle {\mathbf 0}| (I\circ\mathcal V^\text{naive}_{k,\bar k})(z,\bar z)\neq \lim_{z,\bar z\to 0}\langle {\mathbf 0}| (I\circ\mathcal V_{k,\bar k})(z,\bar z)\,,
\ee
because $C_{k,\bar k}$ produces a phase upon moving to the left past $\mathcal V^\text{naive}_{k,\bar k}$. Therefore, definition \eqref{reflector:def} of the reflector state/BPZ conjugate is ambiguous, since the state \eqref{tachyonstate} does not carry any information on which of the two vertex operators ($\mathcal V^\text{naive}_{k,\bar k}$ versus $\mathcal V_{k,\bar k}$) was used to create it. (In fact the ambiguity is present for states with $\mathrm{p}_a\mathrm{w}^a=1\mod 2$.)

We fix the ambiguity by simultaneously specifying the triple of: the reflector state $R$, a choice of $C_{k,\bar k}$ and vertex operator creating any state of the form \eqref{tachyonstate}, and the arbitrary phase parameter $\varphi$ in the conformal inversion \eqref{inversion:maintext} defining the reflector:
\be
R\,,\quad \mathcal V_{k,\bar k}(z,\bar z)= \mathcal V^\text{naive}_{k,\bar k}(z,\bar z)C_{k,\bar k}\,,\quad I(z)=-e^{i\varphi}z^{-1}\,.
\ee
These are subject to the following requirements:
\begin{enumerate}
	\item \emph{Mutual locality:} operators commute at equal (Euclidean) times $|z|=|w|$
	\be
  \label{mutuallocality}
\mathcal V_{k,\bar k}(z,\bar z) \mathcal V_{\ell,\bar \ell}(w,\bar w)|_{|z|=|w|+\varepsilon}= \mathcal V_{\ell,\bar \ell}(w,\bar w) \mathcal V_{k,\bar k}(z,\bar z)|_{|z|=|w|-\varepsilon}\,.
\ee
\item \emph{Covariance under global conformal transformations:} under $z\to  I(z)$,
\be
\label{covariancerequirement}
(I\circ\mathcal V_{k,\bar k})(z,\bar z)=(e^{i\varphi} z^{-2})^h(e^{-i\varphi}\bar z^{-2})^{\bar h}\mathcal V_{k,\bar k}(I(z),I(\bar z))\, ,
\ee
where $h=\tfrac{1}{2} G^{ab}k_a k_b$, $\bar h=\tfrac{1}{2}G^{ab}\bar k_a \bar k_b$ are the conformal weights of $\mathcal V_{k,\bar k}$\,.
\item \emph{Compatibility of the reflector with inversion:}
\be
\label{compatibilityrequirement}
I\circ\cO=\cO^T\,.
\ee
\end{enumerate}
Given the specific choices for $C_{k,\bar k}$ \eqref{cocycleoperators:app} known to satisfy requirement 1, we derive explicit formulas \eqref{transposeassignment} and \eqref{varphitranspositionrule:modes} for BPZ conjugation in the remainder of this section, so  all three requirements are satisfied.

\paragraph{Remarks:}
\begin{itemize}
  \item Requirement 1 can be satisfied in two equivalent ways. If, as is conventional, the position and dual position zero-mode operators commute, then we need $C_{k,\bar k}\neq 1$ to cure the non-commutativity \eqref{badcommutation} of vertex operators. The second possibility is to have \emph{non-commutative} positions, which allows $C_{k,\bar k}=1$ \cite{Sakamoto:1993bc,Lizzi:1997xe,Freidel:2017wst}; in effect one has absorbed $C_{k,\bar k}$ into a redefinition of the position zero-modes. Our derivation of the BPZ conjugation formula \eqref{transposeassignment} is agnostic with respect to which of the two possibilities is realised. However, when we check consistency of BPZ conjugation in Appendix \ref{appendix:consistency}, we need to consider each possibility separately.
  \item Requirement 2 \eqref{covariancerequirement} follows from the definition of the transformation law for a primary operator of weights $h,\bar h$. We mention it separately here for the following reason: when inversion is implemented via a reflector state per requirement 3 \eqref{compatibilityrequirement}, the transformation of e.g.~the naive vertex operator $\mathcal V^\text{naive}_{k,\bar k}$ \emph{is not} of the form \eqref{covariancerequirement}. The same is true for any $\mathcal V_{k,\bar k}$ corresponding to a different choice of cocycle operator $C_{k,\bar k}$. The reason is the explicit appearance of $C_{k,\bar k}$ in the BPZ conjugation rule \eqref{transposeassignment}. 
  \item Although we only discuss the  states \eqref{tachyonstate}, this suffices to completely specify the reflector. This is easily seen in the oscillator picture: since all oscillators arise in the mode expansion of conformal primary operators, their transformation rule under conformal inversion is fixed. Therefore, only the transformation of the states \eqref{tachyonstate} is ambiguous.
  \item Since the choice of reflector state is correlated with the choice of cocycle operator $C_{k,\bar k}$, and the reflector state as well as the state-operator correspondence is employed extensively in the construction of the string field interaction vertices, we conclude we should be using $\mathcal V_{k,\bar k}(z,\bar z)= \mathcal V^\text{naive}_{k,\bar k}(z,\bar z)C_{k,\bar k}$ in the construction of the vertices. This introduces cocycle sign factors in two ways: firstly through $R$ (as we will see shortly), and secondly through $\mathcal V_{k,\bar k}$ which appears inside the CFT correlator that defines the string vertices in the formalism of Zwiebach \cite{Zwiebach:1992ie}. It is natural to conjecture that these are the insertions required for consistency with gauge invariance to all orders.
\end{itemize}

We display the result for the reflector state for the choices
\be
\label{freidelcocycleoperator}
\varphi=0\,,\quad  C_{k,\bar k}\equiv\exp(i\tfrac{1}{2}\pi(k-\bar k)(\hat p+\hat {\bar p}))\,.
\ee
(This is the cocycle operator employed in \cite{Freidel:2017wst}.)
The BPZ conjugate of the state \eqref{nottachyons} is
\be
\left( e^{ikx}e^{i\bar k \bar x} |{\bf 0}\rangle\right)^T=(-1)^{(k^2-\bar k^2)/2}\langle{\bf 0}| e^{ikx}e^{i\bar k \bar x}\,,
\ee
or in terms of positions $\mathrm x^i\equiv x^i+\bar x^i$, dual positions $\tilde {\mathrm x}_i=G_{ij}(\bar x^j-x^j)/2$, momenta, and windings (see \eqref{kbarkintermsofintegers}),
\be
(|{\bf 0}; \mathrm p,\mathrm w\rangle)^T=(-1)^{\mathrm p_a \mathrm w^a} \langle{\bf 0};-\mathrm p,-\mathrm w|\,.
\ee
The inversion \eqref{inversion:maintext} along with the known transformation properties of the conformal primaries $i\pd X(z,\bar z), b(z),c(z)$ implies the following transformation properties for the oscillators appearing in their respective mode expansions:
\be
(\alpha_n^i)^T=(-1)^{n+1}\alpha_{-n}^i\,, \quad (b_n)^T=(-1)^n b_{-n}\,,\quad (c_n)^T=(-1)^{n+1}c_{-n}\,.
\ee
Using \eqref{BPZtransposerules} we define the reflector state $R$ --- and thereby the \lf-algebra cyclic inner product $\kappa$ --- on the entire space. They agree with the reflector state of Kugo and Zwiebach \cite{Kugo:1992md}, obtained in HIKKO string field theory.

\subsection{Construction of the reflector state for the torus CFT}

We now examine possible reflector states for the torus CFT via their equivalent presentation in terms of BPZ conjugation maps.

\paragraph{General features of BPZ conjugation.} We will be defining BPZ conjugation for the entire class of inversion maps depending on a phase $\exp(i\varphi)$:
\be
\label{Iphidef}
I(z)\equiv -e^{i\varphi}z^{-1}\,,\qquad {I}(\bar z)=- e^{-i\varphi} \bar z^{-1}\,.
\ee
The point of allowing a variable phase is that it suggests how to proceed in order to avoid branch cuts associated with expressions like $(-1)^h$. This transformation is involutive for arbitrary values of $\varphi$:
\be
 I^2(z)= z\,.
\ee

A conformal primary $\mathcal O(z,\bar z)$ of weights $h,\bar h$ must transform such that
\be
\begin{split}
({I}\circ\mathcal O)({I}(z),{I}(\bar z)) (d {I}(z))^h (d {I}(\bar z))^{\bar h}= \mathcal O(z,\bar z) dz^h d\bar z^{\bar h}\,,\\
\iff ({I}\circ\mathcal O)({I}(z),{I}(\bar z)) (e^{i\varphi} z^{-2})^h(e^{-i\varphi}\bar z^{-2})^{\bar h}=\mathcal O(z,\bar z)\,.
\end{split}
\ee
We will be assuming that the weights are real and satisfy the integrality condition
\be
\bar h- h\in\mathbb Z\,,
\ee
which is always true for the torus CFT. This will ensure the absence of branch cuts. For example,
\be
(e^{i\varphi} z^{-2})^h(e^{-i\varphi}\bar z^{-2})^{\bar h}=\exp(-is\varphi)|z|^{-2\Delta} (z/\bar z)^s\,,
\ee
where we wrote $h,\bar h$ in terms of scale dimension $\Delta$ and spin $s$:
\be
\label{DeltaSdefinition}
\Delta=h+\bar h\in\mathbb R\,,\qquad s=\bar h-h\in\mathbb Z\,.
\ee
This leads to a sensible formula for the transformation of a conformal primary $\mathcal O$:
\be
\label{}
(I\circ\mathcal O)(z,\bar z)=(e^{i\varphi} z^{-2})^h(e^{-i\varphi}\bar z^{-2})^{\bar h}\mathcal O(I(z),I(\bar z)) =e^{-is\varphi}|z|^{-2\Delta} (z/\bar z)^s \mathcal O(I(z),I(\bar z))\,.
\ee

Given the mode expansion (valid in the Neveu-Schwarz sector)
\be
\mathcal O(z,\bar z)=\sum_{n,\bar n \in\mathbb Z}\frac{\mathcal O_{n,\bar n}}{z^{n+h}{\bar z}^{\bar n+\bar h}}\,,
\ee
we deduce the transformation of the modes:
\be
\label{varphitranspositionrule:modes}
{I}\circ \mathcal O_{n,\bar n}=(-1)^{n+\bar n} e^{i\varphi(n-\bar n)} (-1)^{(\bar h -h)} \mathcal O_{-n,-\bar n}\,.
\ee
This requires writing $-e^{-i\varphi}=e^{-i(\varphi+\pi)}$ inside ${I}(z)$ in the mode expansion:
\be
(-e^{i\varphi})^{n+h}(-e^{-i\varphi})^{\bar n+\bar h}=e^{i(n+ h)(\varphi+\pi)}e^{-i(\varphi+\pi)(\bar n+\bar h)}
\ee 
Notwithstanding the presence of $\varphi$, \eqref{varphitranspositionrule:modes} differs from the literature in the factor $(-1)^{\bar h-h}$ (instead of $(-1)^{h+\bar h}$). Since $\bar h-h$ is an integer, there is no branch cut.

\paragraph{BPZ conjugation of the state \eqref{nottachyons}.} The ``naive'' vertex operator reads
\be
\label{naive:def:app}
\mathcal V^\text{naive}_{k,\bar k}(z,\bar z)\equiv:\exp(ik_i \hat X(z)):\; :\exp(i\bar k_i \hat {\bar X}(\bar z)):\,.
\ee
Conformal normal ordering is actually ambiguous as far as this expression is concerned, as it does not specify the ordering of position zero-modes with respect to momentum zero-modes $\alpha_0$, $\bar \alpha_0$. (This was also pointed out in \cite{Freidel:2017wst}.) Resolving the ambiguity by declaring the momentum zero-modes to be annihilation operators leads to the explicit expression
\be
\begin{split}
:\exp(ik_i \hat X(z)):\equiv\mathcal V^-_k(z)\exp(i k_i x^i) z^{k_i\alpha^i_0} \mathcal V^+_k(z)\\
\mathcal V^-_k(z)=\exp\left(\sum_{n<0}\frac{-1}{n} \frac{k_i\alpha_n^i}{z^n}\right)\,,\qquad \mathcal V^+_k(z)\equiv \exp\left(\sum_{n>0}\frac{-1}{n} \frac{k_i\alpha_n^i}{z^n}\right)
\end{split}
\ee
and the analogous expression for the antiholomorphic operator. We will henceforth set the non-compact momenta to zero, omit indices and factors of $G_{ij}$, and write $\alpha_0=p$, $\bar \alpha_0=\bar p$ (where the conjugate momenta $p,\bar p$ to $x,\bar x$ were defined in \eqref{leftrightmomenta}).

Since $i\pd X(z),i\pd \bar X(\bar z)$ are conformal primaries of weights $h=1$, $\bar h=1$ respectively, \eqref{varphitranspositionrule:modes} yields
\be
\label{bpzmatteroscillators}
{I}\circ\alpha_n=-(-e^{i\varphi})^n \alpha_{-n}\,,\quad {I}\circ\bar \alpha_n=-(-e^{-i\varphi})^n \bar\alpha_{-n}\,.
\ee
In particular
\be
\alpha_0^T=-\alpha_0
\ee
for all $\varphi$, which implies
\be
p^T=-p\,,\quad \bar p^T=-\bar p
\ee
for the (anti)holomorphic momenta \eqref{leftrightmomenta}.

We now proceed to solve the requirements \eqref{covariancerequirement} and \eqref{compatibilityrequirement} for the vertex operator $\mathcal V_{k,\bar k}$ that includes the following $C_{k,\bar k}$ insertion (which we leave arbitrary for the moment)
\be
\mathcal V_{k,\bar k}(z,\bar z)\equiv \mathcal V^\text{naive}_{k,\bar k}(z,\bar z) C_{k,\bar k}\,.
\ee
Together, they are equivalent to the following equality: 
\be
\label{primaryrequirement}
\Big(\mathcal V_{k,\bar k}(z,\bar z)\Big)^T=(e^{i\varphi} z^{-2})^h(e^{-i\varphi}\bar z^{-2})^{\bar h}\mathcal O({I}(z),{I}(\bar z))\,,
\ee
for weights
\be
\label{tachyonweights}
h=\tfrac{1}{2} G^{ab}k_a k_b\,,\qquad \bar h=\tfrac{1}{2}G^{ab}\bar k_a \bar k_b\,.
\ee
Formula \eqref{kbarkintermsofintegers} implies $\bar h-h$ is indeed an integer:
\be
\bar h-h=\mathrm p_a\mathrm w^a\,.
\ee

Given the transformation of the matter oscillators \eqref{bpzmatteroscillators}, we easily find that the non-zero mode contributions transform as desired:
\be
\label{bpznonzeromodes}
\big(\mathcal V^{\pm}_k(z)\big)^T=\mathcal V^\mp_k({I}(z))\,.
\ee
If we abbreviate
\be
{I}(z)=z'\,,\quad {I}(\bar z)=\bar z'\,,
\ee
then $\mathcal V_{k,\bar k}({I}(z),{I}(\bar z))$ reads
\be
\mathcal V _{k,\bar k}(z',\bar z')=\bar{\mathcal V}^-_{\bar k}(\bar z')\mathcal V^-_k(z')\;  e^{ikx}(z')^{kp} e^{i\bar k \bar x}(\bar z')^{\bar k \bar p} C_{k,\bar k} \; \mathcal V^+_k(z')\bar{\mathcal V}^+_{\bar k}(\bar z') 
\ee
while the left-hand side of \eqref{primaryrequirement} is (using the transposition rules \eqref{BPZtransposerules} and \eqref{bpznonzeromodes})
\be
\big(\mathcal V _{k,\bar k}(z,\bar z)\big)^T=\bar{\mathcal V}^-_{\bar k}(\bar z')\mathcal V^-_k(z')\; \Big( e^{ikx} z^{kp} e^{i\bar k \bar x} \bar z^{\bar k \bar p} C_{k,\bar k}\Big)^T \; \mathcal V^+_k(z')\bar{\mathcal V}^+_{\bar k}(\bar z')\,.
\ee
We moved $C_{k,\bar k}$ past the non-zero mode contributions in both formulas, which is legal because it only depends on the operators $p,\bar p$. We find condition \eqref{primaryrequirement} is equivalent to a condition on zero-modes alone:
\be
\label{primality2}
\Big(e^{ikx} z^{kp} e^{i\bar k \bar x} \bar z^{\bar k \bar p} C_{k,\bar k}\Big)^T= (-z'/z)^h(-\bar z'/\bar z)^{\bar h} e^{ikx}(z')^{kp} e^{i\bar k \bar x}(\bar z')^{\bar k \bar p} C_{k,\bar k}\,.
\ee
We proceed to solve this for $(e^{ikx}e^{i\bar k \bar x})^T$.

The left-hand side is
\be
\Big(e^{ikx} z^{kp} e^{i\bar k \bar x} \bar z^{\bar k \bar p} C_{k,\bar k}\Big)^T= C^T_{k,\bar k}  \bar z^{-\bar k \bar p} (e^{i\bar k \bar x})^T z^{-kp} (e^{ikx})^T\,.
\ee
Since $[x,\bar p]=0\implies [x^T,\bar p]=0$, we can act with $z^{kp}\bar z^{\bar k \bar p}$ from the left and get
\be
C^T_{k,\bar k} (e^{i\bar k \bar x})^T(e^{ikx})^T \,.
\ee
On the right-hand side of \eqref{primality2}, acting with $z^{kp}\bar z^{\bar k \bar p}$ from the left leads to terms
\be
 z^{kp} e^{ikx} z^{-kp}= z^{k^2} e^{ikx}= z^{2h} e^{ikx}
 \ee
 and the analogous ones for the barred sector (for $h,\bar h$ of \eqref{tachyonweights}). Writing $-e^{i\varphi}=e^{i(\varphi+\pi)}$ to manage objects raised to $h,\bar h$, we find
\be
 z^{kp}\bar z^{\bar k \bar p}(-z'/z)^h(-\bar z'/\bar z)^{\bar h} e^{ikx}(z')^{kp} e^{i\bar k \bar x}(\bar z')^{\bar k \bar p} C_{k,\bar k}=e^{i\varphi(\bar h-h)}e^{i(\varphi+\pi)(kp-\bar k\bar p)} e^{ikx} e^{i\bar k \bar x} C_{k,\bar k}\,.
\ee
Rewriting the operator in the exponent using \eqref{kbarkintermsofintegers} and \eqref{matterzeromodesintermsofwindingandmomentum:appendix},
\be
kp-\bar k\bar p=-(\mathrm p_i \hat {\mathrm w}^i + {\mathrm w}^i \hat {\mathrm p}_i)\,,
\ee
we see it takes values in $\mathbb Z$, which shows the expression is well-defined.

Altogether, we have found that the requirement \eqref{primaryrequirement} is equivalent to
\be
\label{transposeassignment}
(e^{ikx}e^{i\bar k \bar x})^T= e^{i\varphi(\bar h-h)} e^{i(\varphi+\pi)(kp-\bar k\bar p)} (C^T_{k,\bar k})^{-1}e^{ikx}e^{i\bar k \bar x} C_{k,\bar k}\,.
\ee
This formula must be true irrespective if the positions are commutative (in which case the cocycles are present) or non-commutative (in which case they can be omitted), because we never moved $x$ past $\bar x$ or $C_{k,\bar k}$ past either.
We explicitly check in appendix \ref{appendix:consistency} that the assignment \eqref{transposeassignment} is consistent in that
\be
((e^{ikx}e^{i\bar k \bar x})^T)^T=e^{ikx}e^{i\bar k \bar x}\,,\quad [x,\bar x]^T=[\bar x^T,x^T] \, .
\ee

\section{Homotopy transfer in closed string field theory}
\label{sec:homotopystrings}

In Sen's work \cite{Sen:2016qap}, it was suggested that \emph{any} projector $P$ satisfying
\be
\label{senconditions}
[P,b_0^+]=[P,L_0^+]=[P,Q_{\rm B}]=0
\ee
could be used to construct an effective theory of string states in the image $\im(P)$ of $P$, whose effective Lagrangian is calculated through a Feynman diagram expansion with (Siegel gauge) propagators involving only states in $\ker(P)=\im(\one-P)$. In this section, we apply the machinery of homotopy transfer to realise this idea. 

As reviewed in \cite{Arvanitakis:2020rrk} {and sections \ref{sec:integratingout} and \ref{sec:zeromodes}}, given a projector $P$ in a generic field theory one must construct a homotopy map $h$ satisfying the relation
\begin{equation}\label{homotopyID}
    P = \one + h \pd + \pd h
\end{equation}
along with the following compatibility conditions with the cyclic structure:
\begin{equation}\label{eq:ortho}
    \kappa(Px_1,(\one-P)x_2)=0\, ,\quad \kappa(hx_1,hx_2)=0 \quad \forall x_1,x_2\in X\, .
\end{equation}
Additionally, technical simplifications occur when $h$ satisfies the `side conditions'
\begin{equation}\label{sideconditions}
    h^2 = hP = Ph = 0
\end{equation}
which can always be assumed to hold (possibly by redefining $h$).

In the closed string field theory context, the $1$-bracket is $\pd = Q_{\rm B}$ and the cyclic inner product is $\kappa(\--,\--)=R(\--,c_0^-\--)$. 
{For the gauge-fixed Euclidean theory, as  in \cite{Sen:2016qap}, the existence of a homotopy map $h$ satisfying \eqref{homotopyID} and \eqref{sideconditions} follows from the conditions \eqref{senconditions}. The additional conditions \eqref{eq:ortho} only have to be imposed when a cyclic structure is present. We discuss the general situation in the next subsections.}

\subsection{Integrating out in string  theory}
\label{sec:stringintegratingout}

{We begin with a few general remarks.} String theory in Minkowski space can be regarded as a field theory with an infinite set of  fields, and integrating out a set of massive fields can be treated in the way outlined in {sections \ref{sec:integratingout} and \ref{sec:zeromodes}}.
For any of the superstring theories in 10-dimensional Minkowski space, there is a massless sector consisting of the relevant supergravity theory  together with an infinite tower of massive fields with masses given by the square root of an integer times the string mass $m_s$.
{The effective Euclidean supergravity theory is obtained by first Wick rotating to Euclidean signature then integrating out all the massive fields and in particular would inherit any $L_\infty $ structure that may be present in the Euclidean superstring theory.}
On Wick-rotating back to Lorentzian signature, the effective supergravity will have poles given by the masses of the fields that are integrated out and the low-energy effective action for which
 $|p^2|<<  m_s ^2$ will be non-singular and will have a derivative expansion, giving the supergravity action plus higher derivative corrections.

For the bosonic string in 26-dimensional Minkowski space, there is a tachyon field in addition to the massless fields and infinite tower of massive fields.
The massive fields can be integrated out as before to give an effective field theory for the massless fields and tachyon.  
Integrating out the tachyon is problematic, however, as there is a zero-mode even in Euclidean space, reflecting the pathology of a theory with a tachyon.

Similar results arise for string theory compactified on a torus, with all massive modes (including Kaluza-Klein modes and winding modes) integrated out.
However, there is a subtlety arising in the heterotic, type I and bosonic string theories as   extra massless fields arise at special points in the torus moduli space as a result of symmetry enhancement. If one constructs the effective field theory at a generic point in moduli space, then it can  have singularities at the points in moduli space at which symmetry enhancement occurs.
An alternative is to keep all modes that become massless at some point in moduli space, as envisaged in \cite{GIVEON1991422,Giveon_1994};  this involves an infinite number of fields, but at any given point in moduli space only a finite number will be massless.

\subsection{Propagator}

{Let us now come back to the homotopy formulation.} The homotopy map satisfying the conditions {\eqref{homotopyID}--\eqref{sideconditions}} is constructed as the propagator for the states in $\ker(P)$. The usual propagator in string field theory for states of physical ghost number is the Siegel gauge propagator \cite{Siegel:1985phi}, often written $G=b_0^+ (L_0+\bar L_0)^{-1}=b_0^+(L_0^+)^{-1}$. 
{After continuing to Euclidean signature, $L_0^+$ has no zero-modes (other than the trivial ones from zero-momentum massless fields, as in section \ref{sec:zeromodesintro}) and is invertible on the non-trivial sector. However, in Lorentzian signature all the physical states
 are in the kernel of  $L_0^+$.}

{
To discuss Lorentzian signature homotopy transfer and deal with the zero-modes of $L_0^+$,} one introduces a projector $\Pi:X\to X$ onto $\ker(L_0^+)$ {as in section \ref{sec:zeromodes}}. Since $L_0^\pm$ is diagonal in the Fock space basis:
\be
\label{L0commutator}
[L_0^+,\phi_n]=-n\phi_n
\ee
for any oscillator $\phi_n$ of mode number $n$, there exists a canonical choice for $\Pi$ which is diagonal in the Fock space basis. We can now write down a well-defined operator $G:X\to X$:
\be
\label{siegelprop}
G\equiv b_0^+ (L_0^+)^{-1}(\one-\Pi)\,,
\ee
i.e.~$G=b_0^+ (L_0^+)^{-1}$ on $(\one-\Pi)X$ and $G=0$ on $\Pi X$.
This Siegel gauge propagator realises a homotopy $X \to \im \Pi = \ker L_0^+$: since
\be
\{b_0^+,Q_{\rm B}\}=L_0^+ \,, \quad (b_0^+)^2 = 0\,, \quad (Q_{\rm B})^2 = 0\, ,
\ee
both $b_0^+$ and $Q_{\rm B}=\pd$ commute with $L_0^+$, and thus also with $\Pi$. Therefore
\be
G\pd+\pd G=(\one-\Pi)\iff \Pi=\one +\pd(-G) +(-G)\pd\,.
\ee
In fact $-G$ and $\Pi$ clearly satisfy the side condition \eqref{sideconditions} and orthogonality conditions \eqref{eq:ortho} (the latter due to \eqref{SFTisotropicsubspaces} and \eqref{SFTkappaadjointL0level}), and this ensures the existence of a \emph{cyclic} \lf-algebra structure on $\ker L_0^+$.

The Siegel gauge propagator of string field theory is subtly different from the propagator of \cite{Arvanitakis:2020rrk,Arvanitakis:2019ald}. There, the propagator realises a homotopy onto the homology $H(X)$ and the obtained \lf-algebra structure is thus \emph{minimal}, i.e.~with a vanishing 1-bracket. The case of homotopy transfer to $H(X)$ is known as the \emph{minimal model theorem}. It has been shown that minimal models of the \lf-algebra are identified with S-matrix elements, see e.g.~\cite{Nutzi:2018vkl,Macrelli:2019afx,Jurco:2019yfd,Arvanitakis:2019ald,Lopez-Arcos:2019hvg}. This is a very general statement shown recently not just for classical but also for quantum field theories \cite{Arvanitakis:2019ald,Jurco:2019yfd}. In the formulation of \cite{Arvanitakis:2019ald} this can be summarised succinctly as the statement that LSZ reduction formulas realise homotopy transfer onto $H(X)$.  An analogous statement is known in string field theory: it has been argued that minimal models of the associated homotopy algebras correspond to string scattering amplitudes \cite{Witten:1992yj,Kajiura:2001ng,kajiura,Munster:2012gy}. (In fact the last statement predates the general QFT ones mentioned above.)

What is different in string field theory for the Siegel gauge propagator is that the space of worldsheet BRST cohomology classes $H(X)$ is identified with a proper subspace of $\ker (L_0^+)$:
\be
\label{brstcohomologyversusKerL0}
\mathfrak{i} H(X) \subset \ker L_0^+
\ee
with $\mathfrak{i}$ the inclusion as in the discussion above \eqref{eq:identifyhomology}. In fact, $\ker(L_0^+)$ contains states which are not BRST-closed, for example at degree zero (i.e.~physical ghost number 2) states that are not annihilated by the Virasoro generators $L_n$ with $n>0$. Since the Siegel gauge propagator provides a homotopy onto $\ker(L_0^+)$, it realises an instance of homotopy transfer that is not the minimal model theorem.  It is argued however in \cite[section 5.2]{Kajiura:2001ng} and \cite[sections 3.2-3.3]{Konopka:2015tta} (for the $A_\infty$ algebra of open (super)strings) that the resulting $A_\infty$-brackets generically vanish for the states with $L_0=0$ but $Q_{\rm B}\neq 0$, and the obtained $A_\infty$-algebras were accordingly denoted ``almost minimal models'' in the latter reference. This point was also discussed more recently in \cite[section 3.1]{Erbin:2020eyc}. For our purposes this subtlety simply amounts to replacing $H(X)$ by $\ker(L_0^+)$ appropriately for the projector $\Pi$ of section \ref{sec:zeromodes}.

\subsection{Homotopy transfer onto fixed level}

{Following \cite{Sen:2016qap}, in this section we consider integrating out all fields except those of levels $N=\ell$ and $\bar N=\bar\ell$ from (super)string theory. (If this were in Minkowski space, level matching would require
 $ \ell =\bar\ell$, but different values are possible for toroidal compactifications.)
 The resulting effective action could be useful in discussing the scattering of massive states \cite{Sen:2016qap}.} 

We start with projectors
\be
\label{Pell}
P_{\ell}\equiv\int_0^{2\pi}\frac{d\theta}{2\pi}\exp(i\theta(N-\ell))\,,\quad \bar P_{\bar\ell}\equiv\int_0^{2\pi}\frac{d\theta}{2\pi}\exp(i\theta(\bar N-\bar\ell))\,,\qquad \ell,\bar\ell\in\mathbb Z\,,
\ee
to levels $N=\ell$ and $\bar N=\bar\ell$. Since $(N-\ell)$ and $(\bar N-\bar \ell)$ are both diagonal with integer eigenvalues, those are indeed projectors satisfying the {properties $P_\ell P_{\ell'} = 0$ for $\ell \neq \ell'$ and} $P_\ell^2=P_\ell$. Separate projectors in the holomorphic and antiholomorphic sectors are necessary because the level-matching condition \eqref{levelmatchingandantighostcondition} does not imply $N=\bar N$ in a toroidal background, as can be seen from \eqref{stringL0pmexpressions}.  By $\{b_0,Q_{\rm B}\}=L_0$, $\{\bar b_0,Q_{\rm B}\}=\bar L_0$, and the formulas \eqref{leveloperators}, $P_{\ell}$, $\bar P_{\bar\ell}$ commute with each other, the propagator $G$, and the $\ker L_0^+$ projector $\Pi$.  We can thus write a {projector $P$ onto equal left and right levels $\ell$:
\be
P=P_\ell \bar P_\ell\,,
\ee
and a candidate homotopy
\be\label{eq:h}
h=-G(\one-P)
\ee
that satisfies the homotopy relation {$\widehat{P} = \one + h \pd + \pd h$} for the projector $\widehat{P}$
\be
\label{SFTprojector}
\widehat{P}=P+ (\one-P)\Pi\,.
\ee

The above is essentially the situation of sec.~3 (c.f.~\eqref{Phat}), with the zero modes of $L_0^+$ replacing the homology of $\pd$ on account of the subtlety discussed around formula \eqref{brstcohomologyversusKerL0}: homotopy transfer with a homotopy of the form \eqref{eq:h} cannot eliminate states in $\ker(L_0^+)$, so $\widehat{P}$ is a projector to states of level $\ell=\bar \ell$ or states in $\ker(L_0^+)$ (and their linear combinations).}
The homotopy \eqref{eq:h} also satisfies the side conditions \eqref{sideconditions}, since $-G$ does:
\begin{align}
h^2 &= G(\one-P)G(\one-P)=G^2(\one-P)=0\,,\\
 h P &= -G(\one-P)\Pi=-G\Pi(\one-P)=0\,, \\
 P h &= - P G(\one-P) - (\one-P)\Pi G(\one-P) = 0\, .
\end{align}
Since $P_\ell$ and $\bar P_{\bar\ell}$ {(and thus $P$)} are {all} compatible with the cyclic inner product $\kappa$ (due to \eqref{SFTkappaadjointL0level}), the orthogonality conditions \eqref{eq:ortho} are also satisfied.

We have thus shown that for $h$, $P$ as above, the assumptions for homotopy transfer are satisfied. Therefore, the constructions of the prequel paper \cite{Arvanitakis:2020rrk} produce a cyclic \lf-algebra structure describing the tree-level (in string coupling) effective action. The effective degrees of freedom are string states satisfying the level-matching conditions \eqref{levelmatchingandantighostcondition}, as well as at least one of the conditions
\begin{itemize}
\item $N=\bar N=\ell$, or
\item $L_0^+=0$.
\end{itemize}
{In particular, we are left with arbitrary on-shell states (i.e.~in $\ker(L_0^+)$).}

{For example, the transferred binary bracket reads, for $x, y \in PX$,
\begin{align}
\bar b_2(x,y) = {\widehat{P}} b_2(x,y) = P b_2 (x,y)+ (\one-{P})\Pi b_2(x,y)\,,
\end{align}
where we see explicitly the appearance of zero-modes in the last term. Similarly, the homotopy transfer formula for the 3-bracket now yields \cite[formula (2.98)]{Arvanitakis:2020rrk}
\begin{align}\label{eq:b3}
\bar b_3(x_1,x_2,x_3)&={\widehat{P}} b_3(x_1,x_2,x_3)\\
&\quad +{\widehat{P}}\Big(\big[h\big[x_1,x_2\big],x_3\big]+ (-1)^{x_2x_3} \big[h\big[x_1,x_3\big], x_2\big]+(-1)^{x_1(x_2+x_3)} \big[h\big[x_2,x_3\big], x_1\big]\Big)\nonumber
\end{align}
for the quartic vertex, in terms of the string field theory brackets $b_2 \equiv [\cdot\,, \,\cdot]$ and $b_3$.

 In Euclidean signature,  $L_0^+$ is essentially invertible so that $\Pi = 0$ and $\widehat{P} = P$ (ignoring trivial zero modes from constant massless fields).
 Then the only properties of the projector $P$ that are needed to prove the homotopy relation \eqref{homotopyID} and side conditions \eqref{sideconditions} for the homotopy $h= - b_0^+ (L_0^+)^{-1}(\one-P)$ are Sen's conditions \eqref{senconditions}. 
  Then, as claimed below \eqref{sideconditions}, the assumptions for homotopy transfer are satisfied
  and we have a 
  cyclic \lf-algebra structure describing the Euclidean tree-level  effective action.

As discussed in \cite{Sen:2016qap}, this construction of an effective action for fixed level becomes problematic beyond tree-level due to tadpoles involving massless fields.
Following Sen's suggestion \cite{Sen:2016qap}, we could instead project onto the states of levels $N=\ell$ and $\bar N=\bar\ell$ and the massless states. This would require modification of the projector $P$ but otherwise the analysis would go through as before for this case.
}

\section{Double field theory from closed string field theory}
\label{sec:DFT}

String theory on the product of a torus and a  Minkowski space can be regarded as a 
theory of an infinite set of {\it doubled} fields depending on both moment and winding \cite{Siegel:1993th,Siegel:1993bj,Hull:2009mi}.
The double field theory action of \cite{Hull:2009mi} describes the classical ($g_\text{string}=0$) dynamics of the string states  \eqref{doublymassless} (with $N=\bar N=1$), which we will call ``doubly-massless''.
It provides an
 an effective action for the corresponding doubled fields.

The level-matching constraint \eqref{levelmatchingandantighostcondition} with \eqref{stringL0pmexpressions}
implies that the 
 torus momenta $\mathrm{p}_a$ and windings $\mathrm{w}^a$ are constrained by the \emph{weak section condition}
\be
\label{weaksection}
\mathrm{p}_a \mathrm{w}^a=0\,.
\ee
The doubled momentum $\mathbb P\equiv(\mathrm{p}_a,\mathrm{w}^a)$ is lightlike in a split-signature metric. This condition \eqref{weaksection} does not necessarily imply that $\mathrm{p}=0$ or $\mathrm{w}=0$, instead allowing truly doubled configurations
with both momentum and winding non-zero.

In \cite{Hull:2009mi} the action for the doubly-massless states was constructed  from string field theory up to cubic order in the string field. After strengthening \eqref{weaksection} to the \emph{strong section condition}, which   sets half the components of $\mathbb P$ to zero, an action for the doubly-massless modes was determined to all orders \cite{Hohm:2010pp,Hohm:2010jy}. However the resulting strongly constrained theory describes $d$-dimensional physics; the ``doubling'' only survives in the form of manifest $O(d,d)$-covariance. The construction of a truly double field theory satisfying only the weak constraint beyond cubic order has so far not been achieved.

Such a truly doubled theory can be constructed by 
homotopy transfer from string field theory. We will work in Euclidean signature in this section, to eliminate most of the $L_0^+$ zero modes. Formula \eqref{stringL0pmexpressions} implies that states with $L_0^+=L_0^-=0$ in Euclidean signature have levels $N\leq 1$, $\bar N\leq 1$. Therefore, the projector
\begin{align}
P=  P_1 \bar P_1 + (P_0 \bar P_1 + P_1 \bar P_0) + P_0 \bar P_0
\end{align}
to those levels satisfies the homotopy relation $P=\one+h\pd + \pd h$ for the homotopy $h= - b_0^+ (L_0^+)^{-1}(\one-P)$ of \eqref{eq:h}, where now $\Pi = 0$ as there are no zero modes in $(\one - P) X$.

The first term projects to doubly-massless states with $N = \bar{N} =1$, and we will write it as
\be
P_{\rm DFT}=P_1\bar P_1 \, .
\ee
The bracketed terms project to states with $N = 0$, $\bar{N} = 1$ or $N = 1$, $\bar{N} = 0$. For generic torus moduli\footnote{In our conventions, for $B=0$, generic metric moduli have $G_{ab} \mathrm{w}^a \mathrm{w}^b \neq 2$ for any integer-valued vector $\mathrm{w}$. The rectangular torus at the self-dual radius is $G_{ab}={\rm diag}(2,2,\dots)$.} these are massive states, which may become massless at the special points in moduli space where gauge enhancement occurs. As discussed in section \ref{sec:stringintegratingout}, these terms can be omitted from $P$, at the price of having singularities in the effective theory at these points in moduli space. Finally, the last term projects on the tachyon states $N=\bar N=0$, which we must keep here as we are in bosonic string theory. Since superstring theories do not have a tachyon, we expect that starting with e.g.~the NS-NS sector of superstring field theory in its \lf-algebra formulation  \cite{Erler:2014eba,Goto:2015pqv} would lead to a truly doubled theory of the doubly-massless modes \emph{alone}, instead. This would yield a truly doubled DFT via homotopy transfer with the projector $P_{\rm DFT}=P_1\bar P_1$ replacing $P$, and the appropriate space of superstring states replacing $X$.

We now check that this Euclidean DFT with extra states agrees to cubic order
with the double field theory of \cite{Hull:2009mi}.
 Cubic order corresponds to binary brackets, so we need to consider the binary bracket of double field theory. The latter was obtained from classical string field theory by projecting its binary bracket $b_2$ onto $N=\bar N=1$ and then integrating out auxiliary fields. It is more convenient to keep the auxiliary fields, which makes no difference because integrating out auxiliary fields is a special instance of homotopy transfer \cite{Jurco:2018sby}. The cubic interaction term in the DFT Lagrangian (before auxiliary fields are integrated out) corresponds to the 2-bracket
\be
b^\text{DFT}_2 (x,y)={P_\text{DFT} b_2(x, y)}\,,
\ee
for $x,y\in P_{\rm DFT}X$ two doubly-massless states.
Homotopy transfer onto ${ {P}}X$ gives instead
\be
\label{binaryDFTtransferredbracket}
\bar b_2(x,y)=P b_2(x,y)=b^\text{DFT}_2 (x,y)+P_{\rm extra}b_2(x,y) \,,
\ee
with $P_{\rm extra} = (P_0 \bar P_1 + P_1 \bar P_0) + P_0 \bar P_0$. {Let us omit the first two terms in $P_{\rm extra}$, as discussed above; this leaves us with the tachyon. At cubic level, it is known that the setting the tachyon to zero is a consistent truncation, as was done in \cite{Hull:2009mi} (see also \cite{Tseytlin:1991bu,Banks:1991sg}). This removes the extra term in \eqref{binaryDFTtransferredbracket} and then the result of homotopy transfer, after truncating the tachyon, agrees with the known (cubic) double field theory Lagrangian. Alternatively, a complete $L_\infty$ formulation of superstring theory would bypass this issue entirely.}

The next order couplings in the Lagrangian are at quartic order, which correspond to 3-brackets $\bar b_3$.  Denoting $b_2(x,y)=[x,y]$, the result of the homotopy transfer formula is, as in \eqref{eq:b3},
\begin{align}
\bar b_3(x_1,x_2,x_3)&={{P}} b_3(x_1,x_2,x_3)\\
&\quad +{{P}}\Big(\big[h\big[x_1,x_2\big],x_3\big]+ (-1)^{x_2x_3} \big[h\big[x_1,x_3\big], x_2\big]+(-1)^{x_1(x_2+x_3)} \big[h\big[x_2,x_3\big], x_1\big]\Big)\,,\nonumber
\end{align}
where
 \be
 h = - G(\one-P) = -b_0^+ (L^+_0)^{-1} (\one-P)
 \ee
  This formula provides the quartic vertex of weakly constrained DFT (with extra states in bosonic string theory) in terms of the string field theory brackets $b_2$ and $b_3$.

\section{Conclusions}

In this paper we have revisited  weakly
constrained double field theory.
Such a theory can be formally constructed by integrating out all massive string modes \textit{except} for
the massive modes on toroidal backgrounds that arise from massless fields in $10$ or
$26$ dimensions. 
It  cannot be thought of as a  Wilsonian low-energy effective
field theory as the modes that are integrated out have masses that are comparable (or sometimes less than) the masses of the degrees of freedom that are kept.
 This  construction was addressed by Sen a few years
ago,
in the context of closed string field theory, and he argued that it is
indeed possible to integrate out all massive modes except for the fields
of a weakly constrained
double field theory \cite{Sen:2016qap}.
In this paper, we confirmed and clarified this conclusion and further
elaborated on
the technical details of this procedure. In particular, we showed how the algebraic structure of the interactions and symmetries of the weakly constrained double field theory arise from those of the original string theory.

Concretely, we used the formulation of classical (tree-level) field theories in terms of $L_{\infty}$-algebras,
in which the integrating out of modes can be interpreted algebraically
as  homotopy transfer.
Importantly, the existence of a homotopy transfer can be established on
the basis of the free field theory alone. 
To this end, we revisited the  free covariant closed
string field theory on toroidal backgrounds.

The definition of  free covariant closed
string field theory on toroidal backgrounds involves some 
subtle technical issues concerning vertex operators, the proper form of
cocycle factors and the reflector state that to our knowledge had not been fully given in the literature. In this paper, we gave a careful treatment giving 
full details of the construction.
The reflector state is the bilinear form on the space of string states that, together with the worldsheet BRST charge, defines the kinetic term. In \lf{} language it is the inner product on the associated \lf-algebra (up to a ghost zero mode insertion; see \eqref{cyclicdef}). It is known that for closed string fields on a torus background, consistency of the interaction vertices and the reflector requires the insertion of cocycle (sign) factors: momentum and winding-dependent signs that are absent on uncompactified backgrounds, and ensure gauge invariance on compactified backgrounds. However the correct insertions have only been determined in certain non-covariant cubic string field theories \cite{Hata:1986mz,Kugo:1992md,Maeno:1989uc}, and a prescription for the
non-polynomial  covariant theory of Zwiebach \cite{Zwiebach:1992ie}  has so far been missing. We provided an unambiguous prescription that determines the cocycle factors in the reflector, which depends sensitively on details of the state-operator map and the commutation relations of vertex operators on torus backgrounds, that are also afflicted by cocycle signs. (These last signs also appear in relatively recent discussions of subtleties in T-duality \cite{Hellerman:2006tx} and non-commutativity \cite{Freidel:2017wst} of closed strings on torus backgrounds.) Our construction of the reflector unambiguously defines the \emph{free} covariant closed string field theory of \cite{Zwiebach:1992ie} on a torus background, and provides correct sign insertions for the interaction vertices.

We have  elaborated on our discussion in \cite{Arvanitakis:2020rrk} of the
interpretation of integrating out
degrees of freedom in terms of homotopy transfer, giving details on the role of
gauge redundancies and zero modes. 
With regard to the latter, we introduced the concept of \emph{consistent \lf{} truncation}, which is a criterion that identifies when a theory can be viewed as a consistent subsector of a bigger theory in a broad sense. 
The degrees of freedom of the theory are a subsector of the degrees of freedom of the bigger theory, but the relation between the interactions is more subtle: the construction can be viewed as providing a consistent embedding of the smaller theory into the bigger one.
Homotopy transfer can be seen as an algorithm that constructs such a truncation given a subset of the original degrees of freedom. However, a theory can be a consistent \lf{} truncation of another even when no homotopy transfer can exist, which is the case when the zero mode sectors are different, as we have seen.

To do the integrations needed to obtain double field theory from closed string field theory, the string field theory is first Wick-rotated to Euclidean space (and gauge-fixed where necessary). This ensures that there are no issues with zero modes.
 There is then a definite procedure of integrating out
the relevant modes  to arrive at a weakly constrained double field
theory in Euclidean signature. The formulas of
homotopy transfer provide an algorithmic procedure to determine,
starting from the interaction vertices of the full closed string field
theory,
the vertices of double field theory to arbitrary order in fields. It would also determine the symmetries of the double field theory from those of the string theory.
Unfortunately, the interaction vertices of closed string field theory
are
too involved for  this to be a straightforward procedure. 
We hope to return to the  explicit
construction of the weakly constrained
double field theory elsewhere.

This gives the algebraic structure of the Euclidean double field theory in terms of that of the Wick-rotated string theory.
The final step is then to analytically continue the double field theory  to Lorentzian signature.
We do not anticipate any obstruction to doing so (at least for the classical theory arising from the tree-level effective action).
We  emphasize that,
in our opinion, the explicit construction of a genuine (weakly
constrained) double field theory would be already a major advance even
 in Euclidean signature. It would be an important first
step towards such a theory in Lorentzian signature.
Indeed
one may also view the Euclidean theory as a subsector of a larger Lorentzian
theory yet  to be constructed.
To see this, imagine that we perform a time/space split of a gravity
theory, for instance as the first step towards obtaining the Hamiltonian
formulation.
The resulting action will have `potential terms' involving purely
spatial derivatives plus terms involving time derivatives.
The potential terms are separately invariant under spatial
diffeomorphisms and can hence  be viewed as defining Euclidean gravity.
Similarly, one may subject any double field theory to a time/space split
(see, e.g., sec.~3 of \cite{Chiaffrino:2020akd}).
The potential terms then take the form of
a Euclidean double field theory that is invariant under spatial
(generalized)  diffeomorphisms.
It is this subsector of a Lorentzian weakly constrained double field
theory that is guaranteed to exist by the homotopy transfer procedure.

While we focus on the double field theory sector of closed string field theory, from the viewpoint of this paper 
there is nothing special about this subsector. 
Much of the algebraic machinery could equally be applied to another subsector, such as that arising from the projection to a fixed level, or more generally to the effective theory obtained from integrating out an arbitrary subsector of a field theory.
It is to be hoped, however,  that  there {\it is} something special about the double field theory subsector,
 given that it is closed under the T-duality group $O(d,d,\mathbb{Z})$. For instance, it could be that  the non-localities arising  are {in some sense} milder
than for a generic subsector. (A weakly constrained double field theory cannot be completely local, however,  since the product of fields 
obeying the level-matching constraint must be projected to be level-matched, which is a non-local operation \cite{Hull:2009mi}.) 
To see that something like this  is  possible, we recall 
that there are interesting cases in which  there are elegant non-Wilsonian  effective field theories  for  modes that do  not constitute a low energy subsector.
Examples  of this arise for
Kaluza-Klein truncations. Consider type IIB supergravity 
on $AdS_5\times S^5$, for which the radius of the $S^5$ is comparable to the $AdS$ radius. There is then  no separation between massless modes and 
modes that are much heavier, yet there is a perfectly local field theory for the lowest multiplet: maximal $SO(6)$ gauged supergravity 
in $D=5$ \cite{Gunaydin:1984qu,Pernici:1985ju}, which is a consistent truncation of type IIB supergravity, as proved in \cite{Hohm:2014qga}. 
Integrating out the massive Kaluza-Klein modes 
that do not belong to the lowest multiplet
gives a   non-local effective field theory, but here there is a perfectly local 
theory for these modes that can be found from a truncation of the original theory but which does not arise as a  low-energy limit.

Another example arises for Scherk-Schwarz reductions. For supergravity compactified on a circle with a twist by a duality symmetry \cite{Dabholkar:2002sy} (so that there is a monodromy on the circle given by a duality transformation), then the Scherk-Schwarz reduction \cite{Scherk:1979zr} gives  a lower-dimensional supergravity that is a truncation of the full Kaluza-Klein reduction with a duality twist. For a geometric twist this can be viewed as a compactification on a twisted torus \cite{Hull:2005hk} which then defines the Kaluza-Klein spectrum. In the simplest examples there are two  mass scales in this, one is the Scherk-Schwarz mass  given by the duality twist and the other is set by the radius of the circle. The truncation keeps modes whose mass is set by  the Scherk-Schwarz scale and truncates out those set by the compactification scale, even though this may be comparable or smaller. It would be interesting to see if something similar to these examples happens for weakly constrained double field theory or to try to understand systematically under which conditions such ‘non-Wilsonian’ effective field theories take a manageable form. These and other questions we leave for future work.

\subsection*{Acknowledgements}

We would like to thank Roberto Bonezzi, Christoph Chiaffrino, Harold Erbin, Carlo Maccaferri, Martin Schnabl, Jakub Vo\v{s}mera and Barton Zwiebach for useful discussions. We would also like to thank Warren Siegel for helpful correspondence. Special thanks to Barton Zwiebach for sharing his unpublished notes concerning cocycle factors.

The  work  of  ASA  is  supported  in  part  by  the  “FWO-Vlaanderen”  through  the  project G006119N and by the Vrije Universiteit Brussel through the Strategic Research Program “High-Energy Physics”. The work of OH is supported by the European Research
  Council (ERC) under the European Union’s Horizon 2020 research and
  innovation programme (grant agreement No 771862). CMH  is supported   by the STFC Consolidated Grants ST/P000762/1 and ST/T000791/1. The work of VL is supported by the European Research Council (ERC) under the European Union’s Horizon 2020 research and innovation programme, grant agreement No. 740209.

\appendix

\section*{Appendix}

\section{Consistency of the transpose}
\label{appendix:consistency}
We now check the somewhat bizarre rule \eqref{transposeassignment} for consistency. Firstly, it follows from \eqref{BPZtransposerules} that $(\mathcal O^T)^T=\mathcal O$. By moving $C_{k,\bar k}$ past momenta again, we calculate
\be
\begin{split}
\Big((e^{ikx}e^{i\bar k \bar x})^T\Big)^T&=e^{2i\varphi({\bar h}-h)} e^{i(\varphi+\pi)(kp-\bar k\bar p)}e^{ikx}e^{i\bar k \bar x}e^{-i(\varphi+\pi)(kp-\bar k\bar p)}\\
&= e^{2i\varphi(\bar h-h)} e^{2i(\varphi+\pi)(h-\bar h)} e^{ikx}e^{i\bar k \bar x}= e^{2i\pi(h-\bar h)}e^{ikx}e^{i\bar k \bar x}=e^{ikx}e^{i\bar k \bar x}
\end{split}
\ee
since $\bar h-h\in\mathbb Z$.

It remains to check consistency with respect to the assumed zero-mode commutation relations
\be
[x,\bar x]=ic\,,\quad [x,p]=i=[\bar x,\bar p]\,,\quad [x,\bar p]=[\bar x,p]=0\,.
\ee
It is actually more convenient to define (for wavenumbers $\ell,\bar \ell$)
\be
U_{\ell,\bar \ell}=e^{i\ell p}e^{i\bar \ell \bar p} \quad (\implies U^T_{\ell,\bar\ell}=U^{-1}_{\ell,\bar\ell})
\ee
and
\be
V_{k,\bar k}= e^{ikx}e^{i\bar k \bar x}\,;
\ee
Formula \eqref{transposeassignment} then specifies $V^T_{k,\bar k}$:
\be
\label{transposeassignmentintermsofV}
V^T_{k,\bar k}= e^{i\varphi(\bar h-h)} e^{i(\varphi+\pi)(kp-\bar k\bar p)} (C^T_{k,\bar k})^{-1}V_{k,\bar k} C_{k,\bar k}\,.
\ee
Due to the position-position commutation relation $[x,\bar x]=ic$, the inverse is
\be
V^{-1}_{\ell,\bar\ell}= e^{ic \ell \bar\ell} V_{-\ell,-\bar\ell}\,.
\ee

To check the position-momentum commutation relations, we calculate
\be
U_{\ell\bar\ell}V_{k\bar k}U_{\ell\bar \ell}^{-1}= e^{i(k\ell+\bar k \bar \ell)} V_{k\bar k}\,.
\ee
Taking the transpose on both sides gives
\be
\big(U_{\ell\bar\ell}V_{k\bar k}U_{\ell\bar \ell}^{-1}\big)^T=e^{i(k\ell+\bar k \bar \ell)} (V_{k\bar k})^T\,,
\ee
which is easily confirmed to be consistent with \eqref{transposeassignment} when $C_{k,\bar k}$ commutes past momenta.

Finally, we need to check consistency of \eqref{transposeassignmentintermsofV} with the position-position commutation relation. For this we calculate
\be
\label{VVbraidingnoncommutative}
V_{\ell\bar\ell} V_{k\bar k}=\exp(ic(\bar k \ell-k\bar \ell)) V_{k\bar k} V_{\ell\bar\ell}\,.
\ee
Taking the transpose leads to
\be
\label{transposeconsistency}
V_{k\bar k}^T V_{\ell\bar\ell}^T=\exp(ic(\bar k \ell-k\bar \ell)) V_{\ell\bar\ell}^T V_{k\bar k}^T\,.
\ee
This check is somewhat more involved. We will need the ``braiding'' identity
\be
\label{Vexppbraid1}
V_{k\bar k}e^{i(\varphi+\pi)(\ell p-\bar\ell\bar p)}= e^{i(\varphi+\pi)(\bar k \bar \ell-k\ell)}\;e^{i(\varphi+\pi)(\ell p-\bar\ell\bar p)}V_{k\bar k}\,.
\ee
To proceed with the check, we will specialise to two cases:
\begin{enumerate}
	\item $c\neq 0$ and $C_{k,\bar k}=1$ (non-commutative positions without cocycle operator), or\\
	\item $c=0$ and the cocycle operator
	\be
	\label{cocycleoperators:app}
	C_{k,\bar k}\equiv\exp(i\tfrac{1}{2}\pi(k\mp\bar k)(\hat p\pm\hat {\bar p}))\,.
	\ee
	For the up sign this is the operator \eqref{freidelcocycleoperator}. For the down sign, this can be seen as the same operator defined in a dual frame.
\end{enumerate}
In both cases, the corrected vertex operators \eqref{correctedvertex:defmaintext} are known to satisfy bosonic equal-time commutation relations \eqref{mutuallocality}.\footnote{ We are grateful to Barton Zwiebach for communicating unpublished notes where the down-sign $C_{k,\bar k}$ is shown to satisfy \eqref{mutuallocality}.} 

\paragraph{Case 1.~(non-commutative positions):} The left-hand side of \eqref{transposeconsistency} is (after writing $h_k=k^2/2,\bar h_{\bar k}=\bar k^2/2$ etc.~for the conformal weights)
\be
e^{i\varphi(\bar k^2-k^2 + \bar \ell^2-\ell^2)/2}\; e^{i(\varphi+\pi)(k p-\bar k\bar p)} V_{k\bar k} e^{i(\varphi+\pi)(\ell p-\bar\ell\bar p)} V_{\ell\bar\ell}\,.
\ee
Moving $V_{k\bar k}$ to the right first gives a phase
\be
e^{-ic(\bar k\ell-k\bar \ell)+ i(\varphi+\pi)(\bar k \bar \ell-k\ell)}\,.
\ee
due to \eqref{Vexppbraid1} and the noncommutativity \eqref{VVbraidingnoncommutative} of $V_{k\bar k}$ and $V_{\ell\bar\ell}$. Therefore we find
\be
e^{-ic(\bar k\ell-k\bar \ell)+ i(\varphi+\pi)(\bar k \bar \ell-k\ell)}e^{i\varphi(\bar k^2-k^2 + \bar \ell^2-\ell^2)/2}\; e^{i(\varphi+\pi)(k p-\bar k\bar p)}e^{i(\varphi+\pi)(\ell p-\bar\ell\bar p)} V_{\ell\bar\ell} V_{k\bar k} \,.
\ee
Moving $e^{i(\varphi+\pi)(k p-\bar k\bar p)}$ to the right past $V_{\ell\bar\ell}$ picks up another phase due to the braiding formula \eqref{Vexppbraid1}, which cancels the $(\varphi+\pi)$ dependent phase. We have thus found that the left-hand side of \eqref{transposeconsistency} is
\be
V_{k\bar k}^T V_{\ell\bar\ell}^T=e^{-ic(\bar k\ell-k\bar \ell)}V_{\ell\bar\ell}^T  V_{k\bar k}^T\,,
\ee
which is consistent with \eqref{transposeconsistency} if
\be
\exp\big(2ic(\bar k\ell-k\bar \ell)\big)=1\,.
\ee
Expressing the wavenumbers $k,\bar k$ and $\ell,\bar \ell$ in terms of integer-valued momenta and windings via \eqref{kbarkintermsofintegers} yields
\be
(\bar k\ell-k\bar \ell)=\mathrm p_\ell\mathrm w_k-{\mathrm p}_k \mathrm w_\ell\,,
\ee
which leads to a quantisation condition on the non-commutativity parameter $c$:
\be
c\in\pi \mathbb Z\,.
\ee
Fortunately, this is consistent with the value $c=\pi$ or
\be
[x,\bar x]=i\pi
\ee
found by Freidel, Leigh, and Minic \cite{Freidel:2017wst} by demanding correct equal-time vertex operator commutation relations \eqref{mutuallocality}.

\paragraph{Case 2.~(commutative positions with cocycle operator)}
Notwithstanding overall pure phases that always commute, \eqref{transposeassignmentintermsofV} implies
\be
V_{k\bar k}^T V_{\ell\bar\ell}^T\propto e^{i(\varphi+\pi)(kp-\bar k \bar p)} C_{k\bar k} V_{k\bar k} C_{k\bar k}\; e^{i(\varphi+\pi)(\ell p-\bar \ell \bar p)} C_{\ell\bar \ell} V_{\ell\bar \ell} C_{\ell\bar \ell}\,
\ee
and consistency of the transpose \eqref{transposeconsistency} holds if this equals $ V_{\ell\bar\ell}^T  V_{k\bar k}^T$. 

We divide the argument into two parts. Firstly we assume
the nontrivial identity
\be
\label{pain}
C_{k\bar k} V_{k\bar k} C_{k\bar k} \; C_{\ell\bar \ell} V_{\ell\bar \ell} C_{\ell\bar \ell}=C_{\ell\bar \ell} V_{\ell\bar \ell} C_{\ell\bar \ell}\;C_{k\bar k} V_{k\bar k} C_{k\bar k}\,.
\ee
If we use \eqref{pain} to commute factors involving $k$ past factors involving $\ell$, the only phases we find are the ones from the braiding formula \eqref{Vexppbraid1} from commuting $V_{k\bar k}$ past $e^{i(\varphi+\pi)(\ell p-\bar \ell \bar p)}$ towards the right, and $V_{\ell\bar\ell}$ past $e^{i(\varphi+\pi)(kp-\bar k \bar p)}$ towards the left. These cancel.

Proving \eqref{pain} involves another braiding formula
\be
\label{cocycleVbraid}
C_{k\bar k} V_{\ell\bar \ell}=\exp\Big(i\pi/2\big( k\mp\bar k\big)(\ell\pm\bar\ell)\Big) V_{\ell\bar \ell}C_{k\bar k}
\ee
and also
\be
C_{k\bar k} C_{\ell\bar \ell}=C_{k+\ell,\bar k+\bar \ell}\,,
\ee
both of which follow due to the explicit form \eqref{cocycleoperators:app} of $C_{k,\bar k}$. In rearranging the left-hand side of \eqref{pain} to resemble the right-hand side, we pick up a phase from moving $C_{k\bar k}$ past $V_{\ell\bar\ell}$ to the right twice, and $C_{\ell\bar \ell}$ past $V_{k\bar k}$ to the left, also twice. Therefore, the overall phase is
\be
\exp\Big(i\pi\big( k\mp\bar k\big)(\ell\pm\bar\ell)\Big) \exp\Big(-i\pi\big( \ell\mp\bar \ell\big)(k\pm\bar k)\Big)=\exp(\mp 2 i\pi(\bar k \ell-k\bar \ell ))\,.
\ee
As in Case 1.,~this phase is $1$ due to the fact $(\bar k \ell-k\bar \ell )$ is an integer. This concludes the proof of \eqref{pain}.

The conclusion is that the transpose assigment \eqref{transposeassignment} is consistent in both cases.

\bibliography{LINFTY_paper2}

\providecommand{\href}[2]{#2}\begingroup\raggedright\begin{thebibliography}{10}

\bibitem{Arvanitakis:2020rrk}
A.~S. Arvanitakis, O.~Hohm, C.~Hull, and V.~Lekeu, {\it {Homotopy Transfer and
  Effective Field Theory I: Tree-level}},
  \href{http://arxiv.org/abs/2007.07942}{{\tt arXiv:2007.07942}}.

\bibitem{Sen:2016qap}
A.~Sen, {\it {Wilsonian Effective Action of Superstring Theory}},  {\em JHEP}
  {\bf 01} (2017) 108, [\href{http://arxiv.org/abs/1609.00459}{{\tt
  arXiv:1609.00459}}].

\bibitem{Hohm:2017pnh}
O.~Hohm and B.~Zwiebach, {\it {$L_{\infty}$ Algebras and Field Theory}},  {\em
  Fortsch. Phys.} {\bf 65} (2017), no.~3-4 1700014,
  [\href{http://arxiv.org/abs/1701.08824}{{\tt arXiv:1701.08824}}].

\bibitem{Erbin:2020eyc}
H.~Erbin, C.~Maccaferri, M.~Schnabl, and J.~Vo\v{s}mera, {\it {Classical
  algebraic structures in string theory effective actions}},
  \href{http://arxiv.org/abs/2006.16270}{{\tt arXiv:2006.16270}}.

\bibitem{Koyama:2020qfb}
D.~Koyama, Y.~Okawa, and N.~Suzuki, {\it {Gauge-invariant operators of open
  bosonic string field theory in the low-energy limit}},
  \href{http://arxiv.org/abs/2006.16710}{{\tt arXiv:2006.16710}}.

\bibitem{Mnev:2006ch}
P.~Mnev, {\it {Notes on simplicial BF theory}},
  \href{http://arxiv.org/abs/hep-th/0610326}{{\tt hep-th/0610326}}.

\bibitem{Cattaneo:2008ph}
A.~S. Cattaneo and P.~Mnev, {\it {Remarks on Chern-Simons invariants}},  {\em
  Commun. Math. Phys.} {\bf 293} (2010) 803--836,
  [\href{http://arxiv.org/abs/0811.2045}{{\tt arXiv:0811.2045}}].

\bibitem{Alexandrov:2007pd}
V.~Alexandrov, D.~Krotov, A.~Losev, and V.~Lysov, {\it {On Pure Spinor
  Superfield Formalism}},  {\em JHEP} {\bf 10} (2007) 074,
  [\href{http://arxiv.org/abs/0705.2191}{{\tt arXiv:0705.2191}}].

\bibitem{Kajiura:2001ng}
H.~Kajiura, {\it {Homotopy algebra morphism and geometry of classical string
  field theory}},  {\em Nucl. Phys.} {\bf B630} (2002) 361--432,
  [\href{http://arxiv.org/abs/hep-th/0112228}{{\tt hep-th/0112228}}].

\bibitem{Kajiura:2003ax}
H.~Kajiura, {\it {Noncommutative homotopy algebras associated with open
  strings}},  {\em Rev. Math. Phys.} {\bf 19} (2007) 1--99,
  [\href{http://arxiv.org/abs/math/0306332}{{\tt math/0306332}}].

\bibitem{Zwiebach:1992ie}
B.~Zwiebach, {\it {Closed string field theory: Quantum action and the B-V
  master equation}},  {\em Nucl. Phys.} {\bf B390} (1993) 33--152,
  [\href{http://arxiv.org/abs/hep-th/9206084}{{\tt hep-th/9206084}}].

\bibitem{Siegel:1993th}
W.~Siegel, {\it {Superspace duality in low-energy superstrings}},  {\em
  Phys.Rev.} {\bf D48} (1993) 2826--2837,
  [\href{http://arxiv.org/abs/hep-th/9305073}{{\tt hep-th/9305073}}].

\bibitem{Siegel:1993bj}
W.~Siegel, {\it {Manifest duality in low-energy superstrings}},  in {\em
  {International Conference on Strings 93}}, 9, 1993.
\newblock \href{http://arxiv.org/abs/hep-th/9308133}{{\tt hep-th/9308133}}.

\bibitem{Tseytlin:1990nb}
A.~A. Tseytlin, {\it {Duality symmetric formulation of string world sheet
  dynamics}},  {\em Phys.Lett.} {\bf B242} (1990) 163--174.

\bibitem{Duff:1989tf}
M.~Duff, {\it {Duality rotations in string theory}},  {\em Nucl.Phys.} {\bf
  B335} (1990) 610.

\bibitem{Hull:2009mi}
C.~Hull and B.~Zwiebach, {\it {Double Field Theory}},  {\em JHEP} {\bf 0909}
  (2009) 099, [\href{http://arxiv.org/abs/0904.4664}{{\tt arXiv:0904.4664}}].

\bibitem{Hata:1986mz}
H.~Hata, K.~Itoh, T.~Kugo, H.~Kunitomo, and K.~Ogawa, {\it {Gauge String Field
  Theory for Torus Compactified Closed String}},  {\em Prog. Theor. Phys.} {\bf
  77} (1987) 443.

\bibitem{Kugo:1992md}
T.~Kugo and B.~Zwiebach, {\it {Target space duality as a symmetry of string
  field theory}},  {\em Prog.Theor.Phys.} {\bf 87} (1992) 801--860,
  [\href{http://arxiv.org/abs/hep-th/9201040}{{\tt hep-th/9201040}}].

\bibitem{Maeno:1989uc}
M.~Maeno and H.~Takano, {\it {Derivation of the Cocycle Factor of Vertex in
  Closed Bosonic String Field Theory on Torus}},  {\em Prog. Theor. Phys.} {\bf
  82} (1989) 829.

\bibitem{Chiaffrino:2020akd}
C.~Chiaffrino, O.~Hohm, and A.~Pinto, {\it {Gauge Invariant Perturbation Theory
  via Homotopy Transfer}},  \href{http://arxiv.org/abs/2012.12249}{{\tt
  arXiv:2012.12249}}.

\bibitem{Batalin:1981jr}
I.~A. Batalin and G.~A. Vilkovisky, {\it {Gauge Algebra and Quantization}},
  {\em Phys. Lett.} {\bf 102B} (1981) 27--31. [,463(1981)].

\bibitem{Batalin:1984jr}
I.~A. Batalin and G.~A. Vilkovisky, {\it {Quantization of Gauge Theories with
  Linearly Dependent Generators}},  {\em Phys. Rev.} {\bf D28} (1983)
  2567--2582. [Erratum: Phys. Rev.D30,508(1984)].

\bibitem{Siegel:1988yz}
W.~Siegel, {\it {Introduction to string field theory}},  {\em Adv. Ser. Math.
  Phys.} {\bf 8} (1988) 1--244,
  [\href{http://arxiv.org/abs/hep-th/0107094}{{\tt hep-th/0107094}}].

\bibitem{doubek2007deformation}
M.~Doubek, M.~Markl, and P.~Zima, {\it Deformation theory (lecture notes)},
  {\em arXiv preprint arXiv:0705.3719} (2007).

\bibitem{Nakahara:2003nw}
M.~Nakahara, {\em {Geometry, topology and physics}}.
\newblock CRC Press, 2003.
\newblock Second edition.

\bibitem{Henneaux:1992ig}
M.~Henneaux and C.~Teitelboim, {\em {Quantization of gauge systems}}.
\newblock Princeton University Press, 1992.

\bibitem{Blumenhagen:2013fgp}
R.~Blumenhagen, D.~Lüst, and S.~Theisen, {\em {Basic concepts of string
  theory}}.
\newblock Theoretical and Mathematical Physics. Springer, Heidelberg, Germany,
  2013.

\bibitem{Freidel:2017wst}
L.~Freidel, R.~G. Leigh, and D.~Minic, {\it {Intrinsic non-commutativity of
  closed string theory}},  {\em JHEP} {\bf 09} (2017) 060,
  [\href{http://arxiv.org/abs/1706.03305}{{\tt arXiv:1706.03305}}].

\bibitem{Lizzi:1997xe}
F.~Lizzi and R.~J. Szabo, {\it {Duality symmetries and noncommutative geometry
  of string space-time}},  {\em Commun. Math. Phys.} {\bf 197} (1998) 667--712,
  [\href{http://arxiv.org/abs/hep-th/9707202}{{\tt hep-th/9707202}}].

\bibitem{Kugo:1989tk}
T.~Kugo and K.~Suehiro, {\it {Nonpolynomial Closed String Field Theory: Action
  and Its Gauge Invariance}},  {\em Nucl. Phys. B} {\bf 337} (1990) 434--466.

\bibitem{Belavin:1984vu}
A.~Belavin, A.~M. Polyakov, and A.~Zamolodchikov, {\it {Infinite Conformal
  Symmetry in Two-Dimensional Quantum Field Theory}},  {\em Nucl. Phys. B} {\bf
  241} (1984) 333--380.

\bibitem{Hata:1985zu}
H.~Hata, K.~Itoh, T.~Kugo, H.~Kunitomo, and K.~Ogawa, {\it {Manifestly
  Covariant Field Theory of Interacting String}},  {\em Phys. Lett. B} {\bf
  172} (1986) 186--194.

\bibitem{Hata:1985tt}
H.~Hata, K.~Itoh, T.~Kugo, H.~Kunitomo, and K.~Ogawa, {\it {Manifestly
  Covariant Field Theory of Interacting String. 2.}},  {\em Phys. Lett. B} {\bf
  172} (1986) 195--203.

\bibitem{Frenkel:1980rn}
I.~Frenkel and V.~Kac, {\it {Basic Representations of Affine Lie Algebras and
  Dual Resonance Models}},  {\em Invent. Math.} {\bf 62} (1980) 23--66.

\bibitem{Goddard:1983at}
P.~Goddard and D.~I. Olive, {\it {Algebras, Lattices, and Strings}},  in {\em
  {Kac-Moody and Virasoro Algebras}}, 11, 1983.

\bibitem{Gross:1985rr}
D.~J. Gross, J.~A. Harvey, E.~J. Martinec, and R.~Rohm, {\it {Heterotic String
  Theory. 2. The Interacting Heterotic String}},  {\em Nucl. Phys. B} {\bf 267}
  (1986) 75--124.

\bibitem{Sakamoto:1989ig}
M.~Sakamoto, {\it {A Physical Interpretation of Cocycle Factors in Vertex
  Operator Representations}},  {\em Phys. Lett. B} {\bf 231} (1989) 258--262.

\bibitem{Erler:1991an}
J.~Erler, D.~Jungnickel, J.~Lauer, and J.~Mas, {\it {String emission from
  twisted sectors: cocycle operators and modular background symmetries}},  {\em
  Annals Phys.} {\bf 217} (1992) 318--363.

\bibitem{Sakamoto:1992ur}
M.~Sakamoto and M.~Tabuse, {\it {The General class of string theories on
  orbifolds}},  \href{http://arxiv.org/abs/hep-th/9202083}{{\tt
  hep-th/9202083}}.

\bibitem{Horiguchi:1992sn}
T.~Horiguchi, M.~Sakamoto, and M.~Tabuse, {\it {Cocycle properties of string
  theories on orbifolds}},  {\em Prog. Theor. Phys. Suppl.} {\bf 110} (1992)
  229--260, [\href{http://arxiv.org/abs/hep-th/9202084}{{\tt hep-th/9202084}}].

\bibitem{Sakamoto:1993bc}
M.~Sakamoto, {\it {Topological aspects of antisymmetric background field on
  orbifolds}},  {\em Nucl. Phys. B} {\bf 414} (1994) 267--298,
  [\href{http://arxiv.org/abs/hep-th/9301054}{{\tt hep-th/9301054}}].

\bibitem{Sakamoto:1994nx}
M.~Sakamoto and M.~Tachibana, {\it {Topological terms in string theory on
  orbifolds}},  {\em Prog. Theor. Phys.} {\bf 93} (1995) 471--482,
  [\href{http://arxiv.org/abs/hep-th/9409123}{{\tt hep-th/9409123}}].

\bibitem{Landi:1998ii}
G.~Landi, F.~Lizzi, and R.~J. Szabo, {\it {String geometry and the
  noncommutative torus}},  {\em Commun. Math. Phys.} {\bf 206} (1999) 603--637,
  [\href{http://arxiv.org/abs/hep-th/9806099}{{\tt hep-th/9806099}}].

\bibitem{Hellerman:2006tx}
S.~Hellerman and J.~Walcher, {\it {Worldsheet CFTs for Flat Monodrofolds}},
  \href{http://arxiv.org/abs/hep-th/0604191}{{\tt hep-th/0604191}}.

\bibitem{GIVEON1991422}
A.~Giveon and M.~Porrati, {\it Duality invariant string algebra and d = 4
  effective actions},  {\em Nuclear Physics B} {\bf 355} (1991), no.~2
  422--454.

\bibitem{Giveon_1994}
A.~Giveon, M.~Porrati, and E.~Rabinovici, {\it Target space duality in string
  theory},  {\em Physics Reports} {\bf 244} (Aug, 1994) 77–202.

\bibitem{Siegel:1985phi}
W.~Siegel, {\it {Covariantly Second Quantized String. 2.}},  {\em Phys. Lett.
  B} {\bf 149} (1984) 157.

\bibitem{Arvanitakis:2019ald}
A.~S. Arvanitakis, {\it {The $L_\infty$-algebra of the S-matrix}},  {\em JHEP}
  {\bf 07} (2019) 115, [\href{http://arxiv.org/abs/1903.05643}{{\tt
  arXiv:1903.05643}}].

\bibitem{Nutzi:2018vkl}
A.~N\"utzi and M.~Reiterer, {\it {Scattering amplitudes in YM and GR as minimal
  model brackets and their recursive characterization}},
  \href{http://arxiv.org/abs/1812.06454}{{\tt arXiv:1812.06454}}.

\bibitem{Macrelli:2019afx}
T.~Macrelli, C.~Sämann, and M.~Wolf, {\it {Scattering Amplitude Recursion
  Relations in BV Quantisable Theories}},
  \href{http://arxiv.org/abs/1903.05713}{{\tt arXiv:1903.05713}}.

\bibitem{Jurco:2019yfd}
B.~Jur\v{c}o, T.~Macrelli, C.~Sämann, and M.~Wolf, {\it {Loop Amplitudes and
  Quantum Homotopy Algebras}},  {\em JHEP} {\bf 07} (2020) 003,
  [\href{http://arxiv.org/abs/1912.06695}{{\tt arXiv:1912.06695}}].

\bibitem{Lopez-Arcos:2019hvg}
C.~Lopez-Arcos and A.~Q. Vélez, {\it {L$_{\infty}$-algebras and the
  perturbiner expansion}},  {\em JHEP} {\bf 11} (2019) 010,
  [\href{http://arxiv.org/abs/1907.12154}{{\tt arXiv:1907.12154}}].

\bibitem{Witten:1992yj}
E.~Witten and B.~Zwiebach, {\it {Algebraic structures and differential geometry
  in 2-D string theory}},  {\em Nucl. Phys. B} {\bf 377} (1992) 55--112,
  [\href{http://arxiv.org/abs/hep-th/9201056}{{\tt hep-th/9201056}}].

\bibitem{kajiura}
H.~{Kajiura}, {\it {Noncommutative Homotopy Algebras Associated with Open
  Strings}},  {\em Reviews in Mathematical Physics} {\bf 19} (Jan., 2007)
  1--99, [\href{http://arxiv.org/abs/math/0306332}{{\tt math/0306332}}].

\bibitem{Munster:2012gy}
K.~M\"unster and I.~Sachs, {\it {Homotopy Classification of Bosonic String
  Field Theory}},  {\em Commun. Math. Phys.} {\bf 330} (2014) 1227--1262,
  [\href{http://arxiv.org/abs/1208.5626}{{\tt arXiv:1208.5626}}].

\bibitem{Konopka:2015tta}
S.~Konopka, {\it {The S-Matrix of superstring field theory}},  {\em JHEP} {\bf
  11} (2015) 187, [\href{http://arxiv.org/abs/1507.08250}{{\tt
  arXiv:1507.08250}}].

\bibitem{Hohm:2010pp}
O.~Hohm, C.~Hull, and B.~Zwiebach, {\it {Generalized metric formulation of
  double field theory}},  {\em JHEP} {\bf 1008} (2010) 008,
  [\href{http://arxiv.org/abs/1006.4823}{{\tt arXiv:1006.4823}}].

\bibitem{Hohm:2010jy}
O.~Hohm, C.~Hull, and B.~Zwiebach, {\it {Background independent action for
  double field theory}},  {\em JHEP} {\bf 1007} (2010) 016,
  [\href{http://arxiv.org/abs/1003.5027}{{\tt arXiv:1003.5027}}].

\bibitem{Erler:2014eba}
T.~Erler, S.~Konopka, and I.~Sachs, {\it {NS-NS Sector of Closed Superstring
  Field Theory}},  {\em JHEP} {\bf 08} (2014) 158,
  [\href{http://arxiv.org/abs/1403.0940}{{\tt arXiv:1403.0940}}].

\bibitem{Goto:2015pqv}
K.~Goto and H.~Matsunaga, {\it {A$_{\infty}$ /L$_{\infty}$ structure and
  alternative action for WZW-like superstring field theory}},  {\em JHEP} {\bf
  01} (2017) 022, [\href{http://arxiv.org/abs/1512.03379}{{\tt
  arXiv:1512.03379}}].

\bibitem{Jurco:2018sby}
B.~Jur\v{c}o, L.~Raspollini, C.~Sämann, and M.~Wolf, {\it {$L_\infty$-Algebras
  of Classical Field Theories and the Batalin-Vilkovisky Formalism}},  {\em
  Fortsch. Phys.} {\bf 67} (2019), no.~7 1900025,
  [\href{http://arxiv.org/abs/1809.09899}{{\tt arXiv:1809.09899}}].

\bibitem{Tseytlin:1991bu}
A.~A. Tseytlin, {\it {On the tachyonic terms in the string effective action}},
  {\em Phys. Lett. B} {\bf 264} (1991) 311--318.

\bibitem{Banks:1991sg}
T.~Banks, {\it {The Tachyon potential in string theory}},  {\em Nucl. Phys. B}
  {\bf 361} (1991) 166--172.

\bibitem{Gunaydin:1984qu}
M.~Gunaydin, L.~J. Romans, and N.~P. Warner, {\it {Gauged N=8 Supergravity in
  Five-Dimensions}},  {\em Phys. Lett. B} {\bf 154} (1985) 268--274.

\bibitem{Pernici:1985ju}
M.~Pernici, K.~Pilch, and P.~van Nieuwenhuizen, {\it {Gauged N=8 D=5
  Supergravity}},  {\em Nucl. Phys. B} {\bf 259} (1985) 460.

\bibitem{Hohm:2014qga}
O.~Hohm and H.~Samtleben, {\it {Consistent Kaluza-Klein Truncations via
  Exceptional Field Theory}},  {\em JHEP} {\bf 01} (2015) 131,
  [\href{http://arxiv.org/abs/1410.8145}{{\tt arXiv:1410.8145}}].

\bibitem{Dabholkar:2002sy}
A.~Dabholkar and C.~Hull, {\it {Duality twists, orbifolds, and fluxes}},  {\em
  JHEP} {\bf 09} (2003) 054, [\href{http://arxiv.org/abs/hep-th/0210209}{{\tt
  hep-th/0210209}}].

\bibitem{Scherk:1979zr}
J.~Scherk and J.~H. Schwarz, {\it {How to Get Masses from Extra Dimensions}},
  {\em Nucl.Phys.} {\bf B153} (1979) 61--88.

\bibitem{Hull:2005hk}
C.~M. Hull and R.~A. Reid-Edwards, {\it {Flux compactifications of string
  theory on twisted tori}},  {\em Fortsch. Phys.} {\bf 57} (2009) 862--894,
  [\href{http://arxiv.org/abs/hep-th/0503114}{{\tt hep-th/0503114}}].

\end{thebibliography}\endgroup

\end{document}